\documentclass[11pt,british]{article}\usepackage[]{graphicx}\usepackage[]{color}
\makeatletter
\def\maxwidth{ %
  \ifdim\Gin@nat@width>\linewidth
    \linewidth
  \else
    \Gin@nat@width
  \fi
}
\makeatother

\definecolor{fgcolor}{rgb}{0.345, 0.345, 0.345}

\usepackage{framed}
\makeatletter
 {\par\unskip\endMakeFramed%
 \at@end@of@kframe}
\makeatother

\definecolor{shadecolor}{rgb}{.97, .97, .97}
\definecolor{messagecolor}{rgb}{0, 0, 0}
\definecolor{warningcolor}{rgb}{1, 0, 1}
\definecolor{errorcolor}{rgb}{1, 0, 0}

\usepackage{alltt}
\usepackage{bm}
\usepackage{geometry}
\usepackage{url}
\usepackage{graphicx}
\usepackage{color,colortbl}
\usepackage{rotating}
\usepackage{tabularx}
\usepackage{comment}
\includecomment{comment}
\usepackage{amssymb,amsmath}
\usepackage{ifxetex,ifluatex}
\usepackage{fixltx2e} 
\usepackage{chngcntr}
\usepackage{changepage} 
\usepackage{endnotes} 
\let\footnote=\endnote

\newcommand{\marked}[1]{{\color{black}#1}}

\usepackage{csquotes}

\usepackage[round,authoryear]{natbib}
\usepackage{babel}

\usepackage{comment}
\usepackage{tipa}
\includecomment{comment}

\IfFileExists{upquote.sty}{\usepackage{upquote}}{}
\ifnum 0\ifxetex 1\fi\ifluatex 1\fi=0 
  \usepackage[utf8]{inputenc}
\else 
  \ifxetex
    \usepackage{mathspec}
    \usepackage{xltxtra,xunicode}
  \else
    \usepackage{fontspec}
  \fi
  \defaultfontfeatures{Mapping=tex-text,Scale=MatchLowercase}
  
\fi

\IfFileExists{upquote.sty}{\usepackage{upquote}}{}
\begin{document}

\begin{center}
  {\bf The Cave of Shadows} \\
  {\bf Addressing the human factor with generalized additive mixed models}
\ \\
\ \\

Harald Baayen$^a$ \\
Shravan Vasishth$^b$ \\
Reinhold Kliegl$^b$ \\ 
Douglas Bates$^c$  \\ \ \\
$^a$ University of T\"{u}bingen, Germany\\
$^b$ University of Potsdam, Germany \\
$^c$ University of Wisconsin-Madison, USA \\
\ \\
\ \\
\date{\today}
\end{center}

\vspace*{10\baselineskip}

\noindent Accepted for publication in {\em Journal of Memory and Language} \\ 

\vspace*{7\baselineskip}

\begin{flushleft}
Corresponding author: \\
R. Harald Baayen \\
Seminar f\"{u}r Sprachwissenschaft \\
Eberhard Karls University T\"{u}bingen \\
Wilhelmstrasse 19 \\
T\"{u}bingen \\
e-mail: harald.baayen@uni-tuebingen.de
\end{flushleft}

\newpage
\vspace*{2\baselineskip}
\begin{center}
  {\bf Abstract}
\end{center}

\noindent
Generalized additive mixed models are introduced as an extension of the
generalized linear mixed model which makes it possible to deal with temporal
autocorrelational structure in experimental data.  This autocorrelational
structure is likely to be a consequence of learning, fatigue, or the ebb and
flow of attention within an experiment (the `human factor').  Unlike molecules
or plots of barley, subjects in psycholinguistic experiments are intelligent
beings that depend for their survival on constant adaptation to their
environment, including the environment of an experiment.  Three data sets
illustrate that the human factor may interact with predictors of interest, both
factorial and metric.  We also show that, especially within the framework of
the generalized additive model, in the nonlinear world, fitting maximally
complex models that take every possible contingency into account is ill-advised
as a modeling strategy.  Alternative modeling strategies are discussed for both
confirmatory and exploratory data analysis.  \\ \ \\

\ \\ \ \\

\begin{flushright}
  All models are wrong, but some are useful. \\
  George Box (1979)
\end{flushright}

\section{Introduction}

Regression models are built on the assumption that the residual errors are
identically and independently distributed.  Mixed models make it possible to
remove one source of non-independence in the errors by means of random-effect
parameters.  For instance, in an experiment with fast and slow subjects, the
inclusion of by-participant random intercepts ensures that the fast subjects
will not have residuals that will tend to be too large, and that the slow
subjects will not have residuals that are too small \citep[see, e.g.][for
detailed examples]{Pinheiro:Bates:2000}.  However, even after including
random-effect parameters in a linear model, errors can still show
non-independence.

For studies on memory and language, it has been known for nearly half a century
that in time series of experimental trials, response variables such as reaction
times elicited at time $t$ may be correlated with earlier reaction times at
$t-k, k \geq 1$
\citep{Broadbent:1971,Welford:1980,Sanders:1998,Taylor:Lupker:2001,Gilden2001,GildenEtAl1995,Baayen:Milin:2010}.
One source of temporal dependencies between trials is the presence of an
autocorrelational process in the errors, potentially representing fluctuations
in attention.  Another source may be habituation to the experiment, possibly in
interaction with decisions made at preceding trials
\citep{masson2013modulation}.  Alternatively, subjects may slow down in the
course of an experiment due to fatigue.  A further source of correlational
structure in sequences of responses is learning.  As shown by
\citet{Marsolek:2008}, the association strengths between visual features and
object names are subject to continuous updating.
\citet{Ramscar:Yarlett:Dye:Denny:Thorpe:2010} and
\citet{Arnon:Ramscar:2012} documented the consequences of within-experiment
learning in the domain of language.  \citet{kleinschmidt2015robust} report
and model continuous updating in auditory processing in the context of
speaker-listener adaptation.  \citet{deVaan:Schreuder:Baayen:2007} reported
lexical decisions at trial $t$ to be co-determined by the lexicality decision
and the reaction time to a prime that occurred previously at $t-40$.
Grammaticality judgements that change in the course of an experiment are
reported by \citet{Dery:Pearson:2015}.  We refer to the ensemble of
learning, familiarization with the task, fatigue, and attentional fluctuations
as adaptive processes, or, in short, the `human factor'.  We also refer to data
in which the human factor plays no role whatsoever as `sterile' data, data that
are not infected in any way by hidden processes unfolding in time series of
experimental trials.

Why might we expect that experimental data are not sterile?  Because, unlike
molecules or plots of barley, \marked{human beings adapt quickly and
continuously} to their environment, \marked{and as the work mentioned above has
shown, this includes the environment} of psycholinguistic experiments.  

When temporal autocorrelations are actually present in the data, but not
brought into the statistical model, the residuals of this model will be
autocorrelated in experimental time.  The proper evaluation of model components
by means of $t$ or $F$ tests presupposes that residual errors are identically
and independently distributed.  By bringing random intercepts and random slopes
into the model specification, clustering in the residuals by item or subject is
avoided.  However, such random slopes and random intercepts do not take care of
potential trial-to-trial autocorrelative structure.  The presence of
autocorrelation in the residuals leads to imprecision in model evaluations and
uncertainty about the validity of any significances reported.  When strong
autocorrelation characterizes the residuals, this uncertainty will make it
impossible to draw well-founded conclusions about statistical significance.

It might be argued that adaptive processes, if present, will have effects that
are so minute that they are effectively undetectable.  If so, the experimental
design, and only the experimental design, could serve as a guide for
determining the statistical model to be fitted to the data.   Alternatively,
one might acknowledge the presence of adaptive processes but claim that their
presence gives rise to random and temporally uncorrelated noise.   Any such
adaptive processes would therefore be expected not to interact with predictors
of theoretical interest.   

However, it is conceivable that adaptive processes are present in a way that is
actually not harmless.  We distinguish two cases.  First, adaptive processes
may be present, without interacting with critical predictors of theoretical
interest.  In this case, measures for dealing with the autocorrelation in the
errors will be required, without however affecting the interpretation of the
predictors.  In this case, elimination of autocorrelation from the errors will
result in p-values that are more trustworthy.  Second, it is in principle
possible that adaptive processes actually do interact with predictors of
theoretical interest in non-trivial ways.  If so, it is not only a potential
autocorrelational process in the residual error that needs to be addressed, but
also and specifically the adaptive processes.  These processes, which
themselves may constitute a considerable source of autocorrelation in the
errors, will need to be examined carefully in order to provide a proper
assessment of how they modulate the effects of the critical predictors.

In this study, we discuss three examples of non-sterile data demonstrably
infected by adaptive processes unfolding in the experimental time series
constituted by the successive experimental trials.  First, we re-analyze a data
set with multiple subjects, and a $2 \times 2 \times 4$ factorial design with
true treatments \citep{Kliegl:Kuschela:Laubrock:2015} and a single stimulus
`item'.    We then consider a mega-study with auditory lexical decision
\citep{ernestus2015baldey} using a regression design with crossed random
effects of subject and item.  The third analysis concerns a self-paced reading
study in which subjects were reading Dutch poems, following up on earlier
analyses presented in \citet{Baayen:Milin:2010}.   

The analyses of these three data sets make use of the generalized additive
mixed model ({\sc gamm}).  Before presenting these analyses, we first provide
an introduction to {\sc gamm}s. Section~4 discusses regression modeling
strategies for dealing with the human factor when conducting confirmatory or
exploratory data analysis, and the final discussion section, after summarizing
the main results, closes with some reflections on the importance of parsimony
in regression modeling.

\section{The generalized additive mixed model}\label{sec:gamms}

In linear regression, a univariate response $y_i$ (where $i$ indexes the
individual data points) is modelled as the sum of a linear predictor $\eta_i$
and a random error term with zero mean.  This linear predictor is assumed to
depend on a set of predictor variables.  Often, the response variable is
assumed to have a normal distribution.  If so, a regression model such as
\[
y_i = \eta_i + \epsilon_i \text{ where } \epsilon_i \underset{\text{ind}}{\sim} N(0,\sigma^2) \text{ and } \eta_i = \beta_0 + \beta_1 x_{1i} + \beta_2 x_{2i}.
\]
describes a response variable $y$ that is modeled as a weighted sum of two
predictors, $x_1$ and $x_2$, together with an intercept ($\beta_0$) and
Gaussian error with standard deviation $\sigma$.

Generalized linear models let the response depend on a smooth monotonic
function of the linear predictor.  This family of models allows the response to
follow not only the normal distribution, but other distributions from
the exponential family, such as Poisson, gamma, or binomial.  An example of a
binomial {\sc glm} with the same linear predictor $\eta$ is
\[
y_i \underset{\text{ind}}{\sim} \text{binom}(\exp(\eta_i)/\{1+ \exp(\eta_i)\},1) 
\text{ where } \eta_i = \beta_0 + \beta_1 x_{1i} + \beta_2 x_{2i}.
\]
This equation specifies that $y_i$ follows a binomial distribution with `number
of trials' = 1, and a probability of success $\exp(\eta_i)/\{1+ \exp(\eta_i)\}$
that is dependent on the predictor variables.  The generalized linear mixed
model ({\sc glmm}) enriches the {\sc glm} with further sources of random noise,
modeled with the help of Gaussian random variables with mean zero and unknown
standard deviation to be estimated from the data.  By way of example, if $y$
denotes response time, $x_1$ the amount of sleep deprivation, and $x_2$
temperature, an experiment carried out with multiple subjects $j$ would be
analyzed with the model

\[
y_{ij} = \eta_{ij} + \epsilon_{ij} 
  \text{ where } \epsilon_{ij} \underset{\text{ind}}{\sim} N(0,\sigma^2) 
  \text{ and } \eta_{ij} = \beta_0 + b_j + \beta_1 x_{1i} + \beta_2 x_{2i}, \text{ with }
  b_j \underset{\text{ind}}{\sim} N(0,\sigma_b^2), 
\]
under the assumption that the only term in the model that has to be adjusted
from subject to subject is the intercept.  In other words, this model assumes
that there are faster and slower subjects, and that in all other respects,
subjects behave in the same way.  Specifically, the effects of the predictors
$x_1$ and $x_2$ are assumed not to vary across subjects.  More complex models
can be obtained by relaxing these assumptions \citep[see,
e.g.,][]{Pinheiro:Bates:2000}.  The $b_j$ given the estimate of $\sigma_b$ are
known as best unbiased linear predictors ({\sc blup}s), conditional modes, or
posterior modes.

A generalized {\em additive} mixed model
\citep{Hastie:Tibshirani:1990,Lin:Zhang:1999,Wood:2006,Wood:2011,wood2015generalized}
is a {\sc glmm} in which part of the linear predictor $\eta$ is itself
specified as a sum of smooth functions of one or more predictor variables.
Thus, a generalized additive (mixed) model is additive in two ways. First, it
inherits from the generalized linear model that the linear predictor is a
weighted sum.  The generalized additive model adds to this functions of one or
more predictors that themselves are weighted sums of basis functions.  An
important property of {\sc gamm}s is that each term in the model specifies a
partial effect, i.e., the effect of that specific term when all other terms in
the model are held constant.  

In what follows, we first discuss univariate splines, illustrated with the time
series of reaction times of one subject (123) in the {\tt KKL} dataset, a
dataset we return to in more detail below.  Following this, we introduce
multivariate splines, using as example lexical decision latencies elicited for
Vietnamese compound words.

\subsection{Univariate splines}

\begin{figure}[h]
  \centering
  \includegraphics[width=\textwidth]{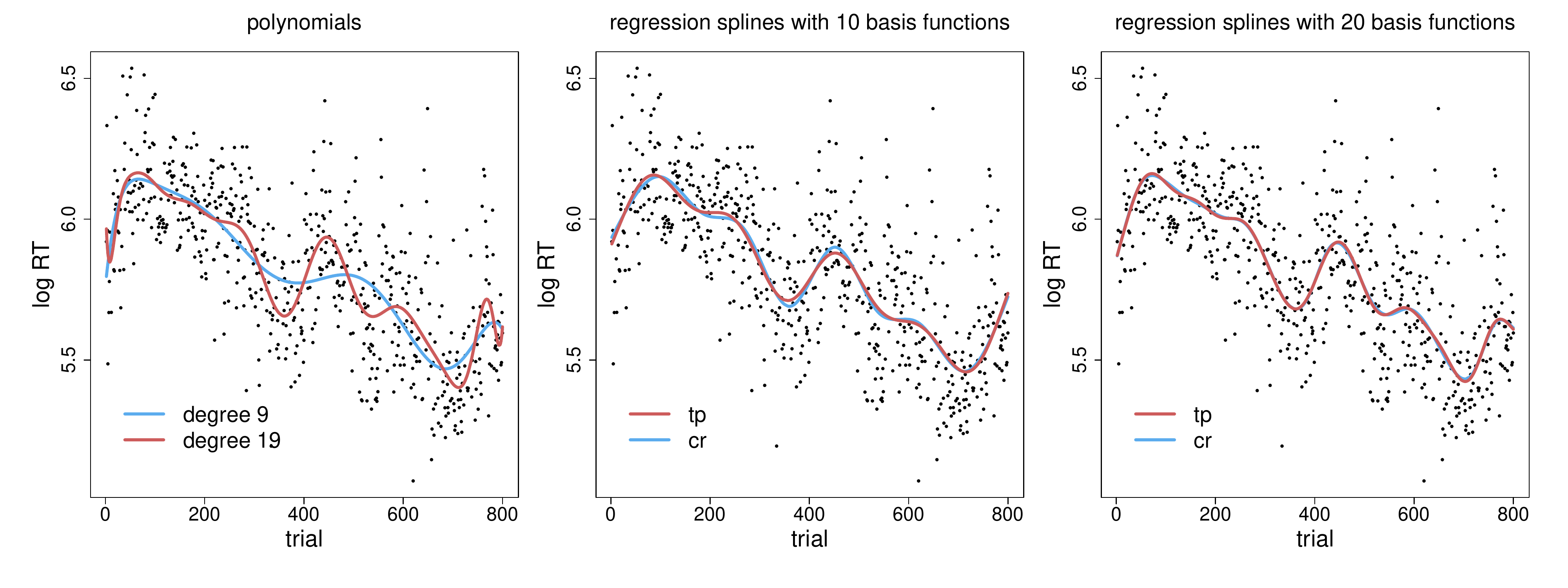}  

  \caption{Log reaction time as a function of trial for subject 123 in the {\tt
  KKL} data set.  Left: fit with orthogonal polynomials of degree 9 and 19 (10
  and 20 parameters including the intercept); Center: fit with regression
  splines with 10 basis functions; Right: fit with regression splines with 20
  basis functions. {\tt cr}: restricted cubic spline; {\tt tp}: thin plate
  regression spline.}

  \label{fig:polyPlus}
\end{figure}

The left panel of Figure~\ref{fig:polyPlus} presents the time series of subject
123 in the {\tt KKL} data set, with trial number (1, 2, \ldots, 800) on the
horizontal axis, and log response time on the vertical axis.  This plot reveals
that as the experiment proceeded, this particular subject tended to respond
more quickly.  Although a linear model fitting a straight line to these data, 
\[
y_i = \beta_0 + \beta_1 x_i +  \epsilon_i,
\text{ where } \epsilon_i \underset{\text{ind}}{\sim} N(0,\sigma^2) 
\]
supports a downward trend ($p < 0.0001$, {\sc aic} =
\ensuremath{-206.46}), it is clear that a straight line does not do
justice to the undulating pattern that appears to ride on top of the linear
trend.  We therefore need to relax the linearity assumption, and allow the
response $y$ to be a smooth function $f$ of $x$:
\[
y_i = \beta_0 + f(x_i) +  \epsilon_i,
\text{ where } \epsilon_i \underset{\text{ind}}{\sim} N(0,\sigma^2). 
\]
The regression smooth $f(x)$ is a weighted sum of a set of $q$ so-called basis
functions defined over the predictor $x$
\citep{Wood:2006,james2013introduction}.  Writing $B_k$ for the $k$-th
basis function, we have that
\begin{equation}
f(x_i) = \sum_{k=1}^{q} B_k(x_i) \beta_k.
\label{eq:basis}
\end{equation}
The question is how to choose these basis functions.  One might consider using
polynomials of $x$, i.e, basis functions of the form
\[
B_k(x_i) = x_i^k, \hspace*{1em} k = 0, 1, \ldots,
\]
which leads to regression models with a polynomial of degree $d$ and $d+1$
parameters:
\[
y_i = \beta_0 x_i^0 + \beta_1 x_i^1 + \beta_2 x_i^2 + \beta_3 x_i^3 + \ldots + \beta_d x_i^d +  \epsilon_i,
\text{ where } \epsilon_i \underset{\text{ind}}{\sim} N(0,\sigma^2). 
\]
The blue curve in the left panel of Figure~\ref{fig:polyPlus} presents the fit
of a polynomial of degree $d = 9$, which requires 10 $\beta$ coefficients  (one
for the intercept, and 9 for the non-zero powers of $x$).  Although this model
provides an improved fit to the data ({\sc aic} = \ensuremath{-274.74}),
visual inspection suggests it oversmoothes the data.  A polynomial of degree
$d=19$, shown in red, follows the trend in the data more closely, and provides
a substantially improved fit ({\sc aic} = \ensuremath{-332.27}).
Unfortunately, the undulations for the earliest and latest trials look
artefactual, and suggest undersmoothing.  More in general, higher-order
polynomials come with several undesirable properties when interest is in the
behavior of the response variable over the full range of the predictor.  The
present artefactual wiggliness at the edges of the predictor domain, where data
are sparse, illustrates one such undesirable property.  Regression splines have
been developed to avoid such artefacts.

\begin{figure}[h]
  \centering
  \includegraphics[width=0.5\textwidth]{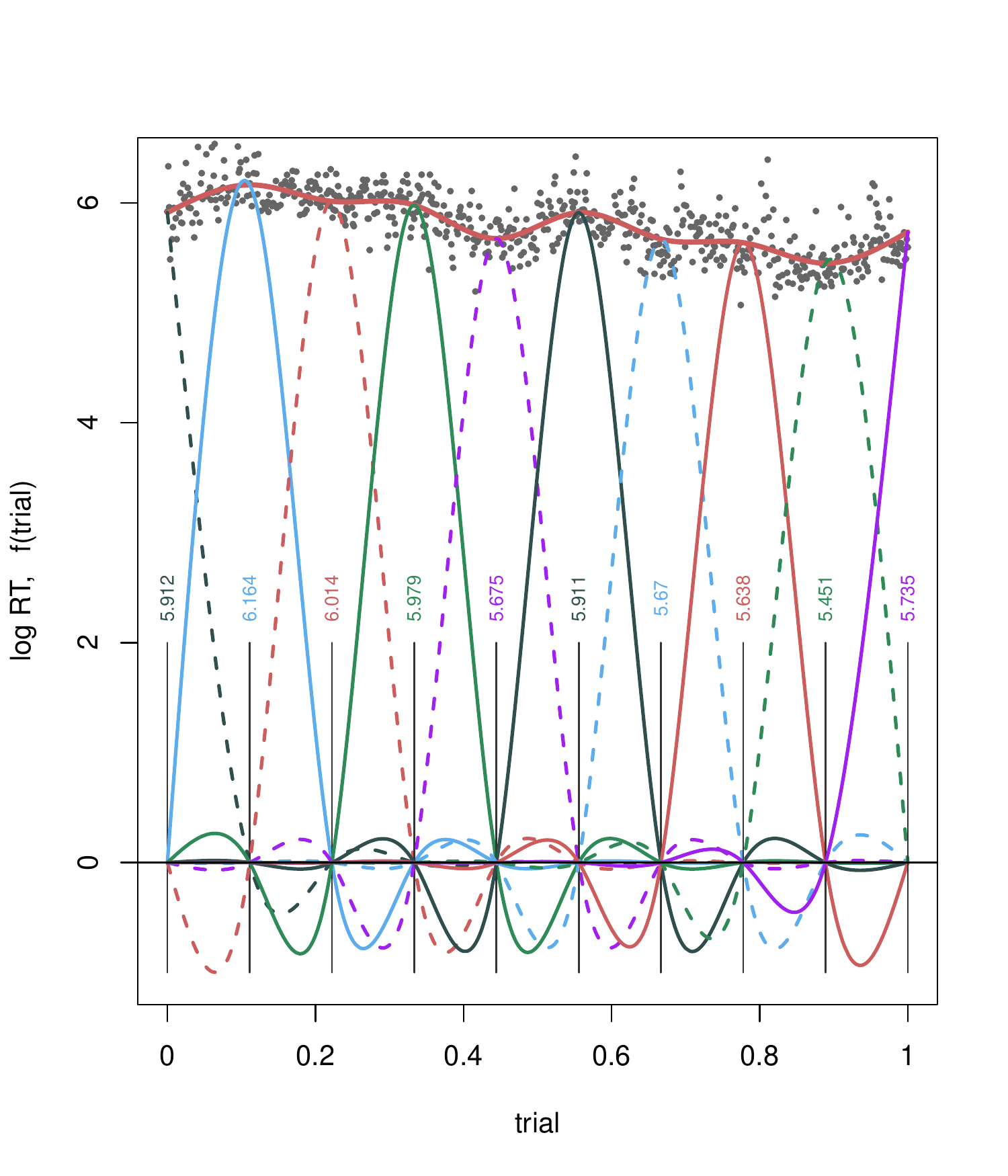}

  \caption{Log reaction time as a function of trial (scaled to the [0,1]
  interval) for subject 123 in the {\tt KKL} data set, with an unpenalized restricted
  cubic spline and its basis functions. Vertical black lines indicate knots.
  The numbers above these lines are the (unpenalized) weights for the basis
  functions.  Note that for any given basis function, the maximum is reached at
  exactly one knot, whereas at all other knots, it is zero.}
  
  \label{fig:cr}
\end{figure}

There are many different kinds of splines, we restrict ourselves here to two
particular splines: restricted cubic splines ({\tt cr}) and thin plate
regression splines ({\tt tp}).  The center and right panels of
Figure~\ref{fig:cr} illustrates these splines for 10 and 20 basis functions
respectively. The blue curves represent restricted cubic splines, and the red
curves, thin plate regression splines.  With 10 basis functions (and 10
parameters), the splines already capture the trend in the data much better than
the corresponding 10-parameter polynomial ({\sc aic} {\tt cr} =
\ensuremath{-316.75}, {\sc aic} {\tt tp} =
\ensuremath{-311.4}), for 20 basis functions, fits are
comparable to that of the polynomial of degree 20 ({\sc aic} {\tt cr} =
\ensuremath{-331.52}, {\sc aic} {\tt tp} =
\ensuremath{-334.06}) but without edge artifacts.

The basis functions for the {\tt cr} regression spline in the center panel of
Figure~\ref{fig:polyPlus} are illustrated in Figure~\ref{fig:cr}.  Again, the
horizontal axis represents trial number, and the vertical axis log response
time. The data points are shown, together with the restricted cubic spline
smooth (in red).  The basis functions all have the same functional form, the
mathematical definition of which can be found in, e.g., Wood (2006, chapter 4).
Each basis function is a curve that itself is made up of sections of cubic
polynomials, under the constraint that the function must be continuous up to
and including the second derivative.  The points at which the sections of the
curve meet are referred to as knots.  In Figure~\ref{fig:polyPlus}, these knots
are indicated by vertical black lines.  The numbers above these black lines
represent the weights $\beta_k$ (cf. equation \ref{eq:basis}) for the basis
functions that have their maximum above these knots.  In this parameterization
of restricted cubic regression splines, any given basis function has its
maximum at one specific knot, and is zero at all other knots.

\begin{figure}
  \centering
  \includegraphics[width=\textwidth]{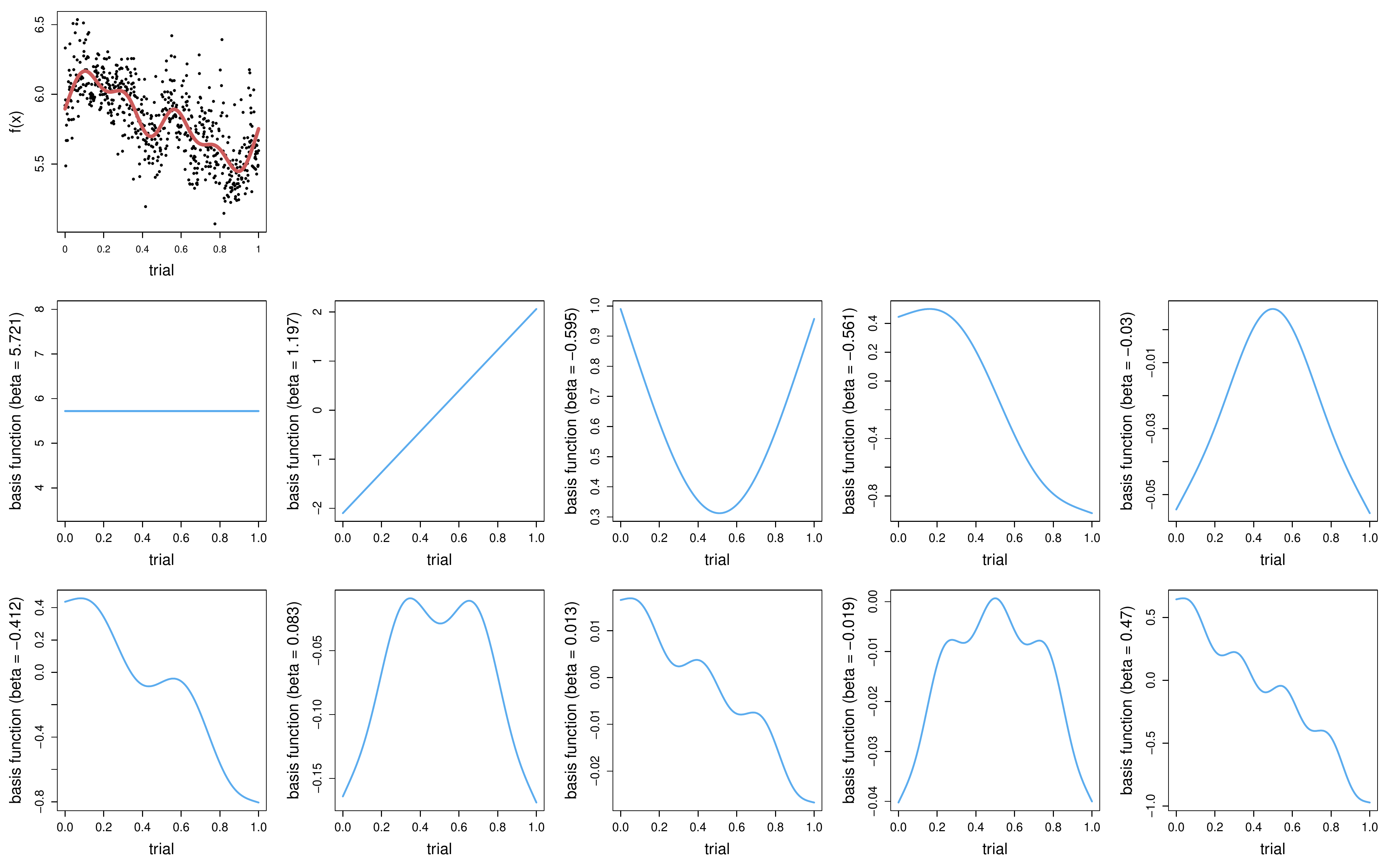}

  \caption{Log reaction time as a function of trial (scaled to the [0,1]
  interval) for subject 123 in the {\tt KKL} data set, with an unpenalized thin plate
  regression spline (upper left) and its basis functions. The Y-axes for the
  basis functions have different scales, as the basis functions have been
  weighted (unpenalized weights in parentheses).  Note, first, that the first
  two basis functions are not wiggly but completely smooth, and that the amount
  of wiggliness of the remaining thin plate basis functions increases from left
  to right and top to bottom.}
  
  \label{fig:tp}
\end{figure}

The basis functions for a thin plate regression spline are constructed in a
different way.  Figure~\ref{fig:tp} illustrates the 10 basis functions for the
{\tt tp} smooth in the second panel of Figure~\ref{fig:polyPlus}.  The first
basis function is a horizontal line, allowing calibration of the intercept. The
second basis function is a straight line, allowing the model to capture linear
trends.  Note that the slope of this line can be reversed by using a negative weight.

Whereas the first two basis functions are completely smooth, the remaining
basis functions are wiggly.  The exact form of these basis functions depends on
the number of basis functions requested, as well as on whether the basis
function is the first, second, third, \ldots, of the requested wiggly basis
functions (for mathematical details, see, e.g., Wood 2006, chapter 4).
What is important is that each successive basis function is more wiggly than the
preceding one.  Thus, each additional basis function makes it possible to model
more subtle aspects of the wiggliness in the data. Here too, negative weights
will reverse the orientation of the basis functions, changing for instance
a parabola that opens upward in a parabola that opens downward.

\begin{figure}[htbp]
  \centering
  \includegraphics[width=0.6\textwidth]{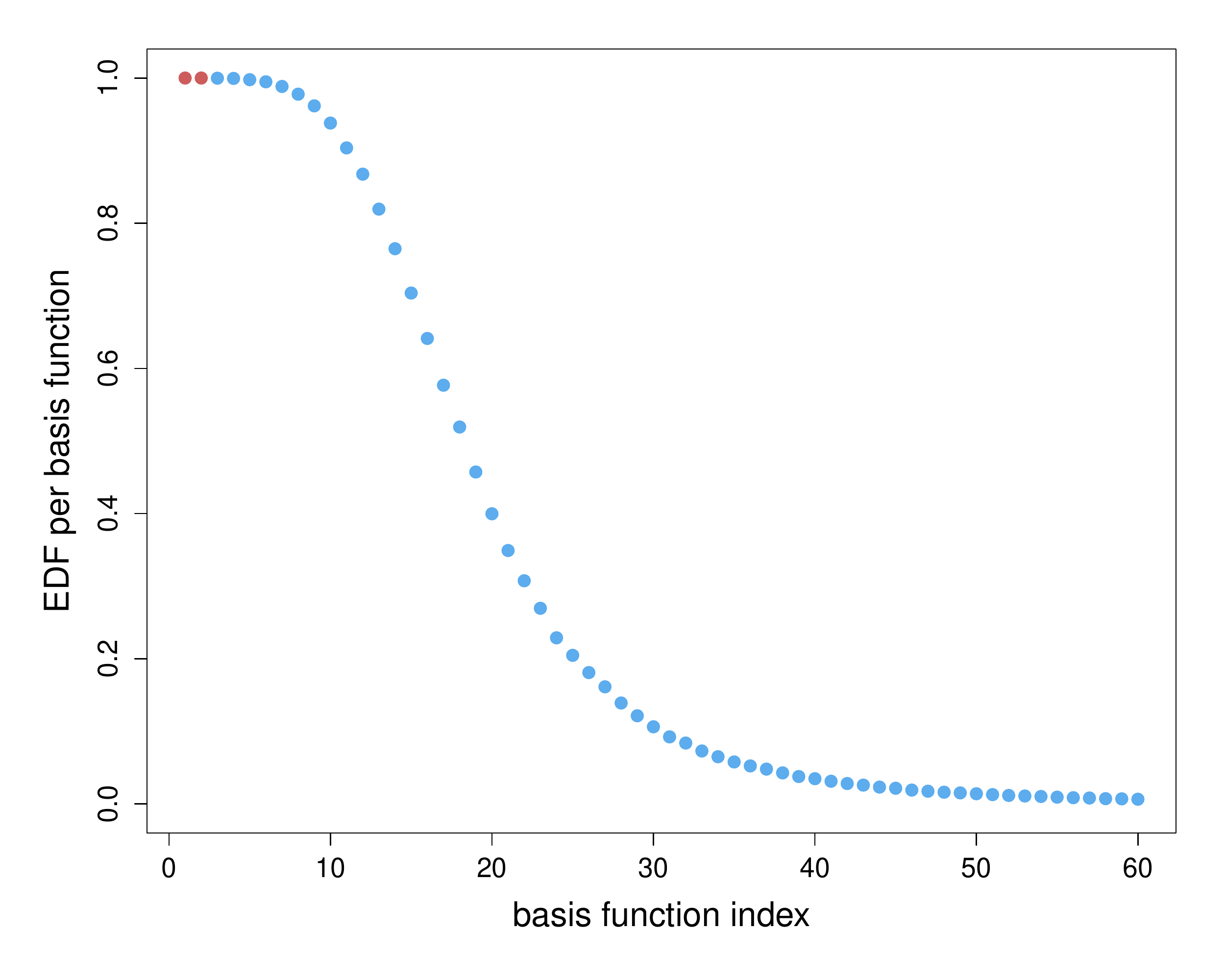}

  \caption{Edf per basis function for a thin plate regression spline with
  penalty order $m=2$ fitted to the time series of reaction times of subject
  123 in the {\tt KKL} data set.  The first two basis functions (in red) are
  completely smooth functions (a horizontal straight line and a tilted straight
  line), which cannot be penalized for wiggliness.}

  \label{fig:edfKKL}
\end{figure}

At this point, we are faced with the question of what the optimal number of
basis functions is.  On one hand, we want to be faithful to the data, but
on the other hand, we also want to avoid overfitting and incorporating
spurious wiggliness, such as observed for the polynomial of degree 20 in
the left panel of Figure~\ref{fig:polyPlus}.  The solution offered by
spline theory is to start with $k$ basis functions, and to select
that vector of estimated coefficients $\hat{\bm{\beta}}$ such that
the quantity Q,
\begin{equation}
Q = \sum_i (y_i - f(x_i))^2 + \lambda \int f^{\prime\prime}(x)^2 dx,
\end{equation}
is minimized.  $Q$ will be smaller when the weights are chosen such that the
summed squared error is smaller.  At the same time, $Q$ will be larger for
smooths with greater wiggliness, quantified by the integral over the squared
second derivative of the smooth.  The balance of the constraint to stay
faithful to the data and to avoid excess wiggliness is regulated by the
smoothing parameter $\lambda$.  When $\lambda = 0$, all that counts is
faithfulness to the data, irrespective of how complex the spline smooth is.  As
$\lambda$ is increased, the complexity of the spline comes into play, and
undersmoothing becomes more and more costly.

The appropriate amount of penalization (given by an optimal $\lambda$) can be
estimated by prediction error methods (cross-validation) or by marginal
likelihood.  The latter method (used in the present study) requires a prior on
the distribution of the coefficients $\bm{\beta}$. This prior expresses
mathematically that the `truth' is more likely to be smooth than wiggly (cf.
Occam's razor).  The smoothing parameter $\lambda$ gets tuned in order that
random draws from the prior on the $\bm{\beta}$ coefficients that is implied by
$\lambda$ have high average likelihood.  This Bayesian approach is also used
for variance estimation, making for easier confidence interval calculation
while at the same time providing good coverage probabilities
\citep{Nychka:1988,Marra:Wood:2012}.\footnote{For a fully Bayesian approach
to generalized additive modeling, see \citet{Wood:2016jagam}, where an
interface between {\tt mgcv} and the {\sc bugs} language is discussed.  With
this interface, it becomes possible to adopt a fully Bayesian approach to
\textsc{GAMM}s.}

Given $k$ basis functions, penalization will typically result in a model with
effective degrees of freedom less than $k$:  $k$ specifies the maximum possible
degrees of freedom for a model term.  However, it is possible that the initial
dimensionality selected is too low.  Doubling the number of basis functions and
re-fitting will show whether this is indeed the case.  Checking that $k$ is not
too restrictive is an essential part of working with {\sc gam}s.

An important consequence of penalization is that the coefficients $\bm{\beta}$
are no longer free to vary. The values of these coefficients will be smaller
than if there were no penalization (i.e., if $\lambda$ were zero).  The extent
to which a coefficient is smaller under penalization, its {\em shrinkage
factor}, is bounded between 0 and 1 and is referred to as its {\em effective
degrees of freedom} ({\tt edf}).  Figure~\ref{fig:edfKKL}\ illustrates the
effective degrees of freedom for a thin plate regression spline with 60 basis
functions, fitted to the time series of reaction times of subject 123 in the
{\tt KKL} dataset. The first two basis functions (in red) are straight lines and
hence receive no penalty for wiggliness.  The remaining basis functions are
wiggly.  Most of the second half of the basis function (index $> 30$) are
severely penalized.  

When a thin plate regression spline is fitted to data with a linear trend, the
wiggly basis functions will be strongly penalized whereas the two linear basis
functions are retained without penalization. Setting aside the {\tt edf} of 1
for the intercept, the {\tt edf}s for such a smooth will be 1 or slightly
greater than 1, as the slope of the second basis function requires 1
parameter.  If the data follow a quadratic trend, the {\tt edf} will be close
to 3, with a strong weight for the third basis function.  However, {\tt edf}s
for thin plate regression splines close to 3 may also be indicative of other
trends, such as downward trends that level off for larger values of the
predictor.

The sum of the {\tt edf}'s of the basis function is used for significance
testing in model comparison.  For instance, a comparison of the abovementioned
models using thin plate regression splines with 10 and 20 basis functions shows
that increasing the number of basis functions results in a decrease of the
residual deviance of 1.1664
at the cost of 15.4248 (the total {\tt edf} of the
more complex model) $-$ 9.6448 (the total {\tt
edf} of the simpler model) $=$
5.78 {\tt edf}s.  An
F-test ($F=6.01, p < 0.0001$) suggests that the investment in a more complex
model pays off.

Thus far, we have considered the time series of only one subject. When multiple
subjects are considered simultaneously, we need a generalized additive {\em
mixed} model ({\sc gamm}). As a first step, we may consider a  model with
by-subject random intercepts $b_j$:
\[
y_{ij} = \beta_0 + f(x_i) + b_j + \epsilon_{ij}
\text{ where } \epsilon_{ij} \underset{\text{ind}}{\sim} N(0,\sigma^2) \text{ and } 
  b_j \underset{\text{ind}}{\sim} N(0,\sigma_b^2).
\]
Random effects are implemented as parametric terms penalized by a ridge penalty
\citep{james2013introduction}, which is equivalent to the assumption that
the coefficients are independently and identically distributed normal random
effects.  The implementation of random effects by means of ridge penalties does
not exploit the sparse structure of many random effects, and hence they are
more costly to compute than corresponding random effects in the linear mixed
model.\footnote{
  In the {\sc gamm} framework, random effects can be thought of as smooths with
  a zero-dimensional null space, i.e., as splines with no completely smooth
  basis functions (such as the first two basis functions in
  Figure~\ref{fig:tp}). At the same time, the prior on wiggliness is equivalent
  to the requirement that the coefficients of the smooth ($\bm{\beta}$) follow
  a multivariate normal distribution with zero mean.  In other words, they can
  be viewed as a source of Gaussian noise, just as the random effects in the
  linear mixed model.
}

The above model is unsatisfactory, however, because it assumes that each
subject goes through the experiment in exactly the same way.  At the very
least, we need to allow for separate regression splines $f_j(x)$ for different
subjects $j$:
\[
y_{ij} = \beta_0 + f_j(x_i) + b_j + \epsilon_{ij}
\text{ where } \epsilon_{ij} \underset{\text{ind}}{\sim} N(0,\sigma^2) \text{ and } 
  b_j \underset{\text{ind}}{\sim} N(0,\sigma_b^2).
\]
This model incorporates a nonlinear interaction of subject by trial, 
but restricts by-subject random effects to the intercept.  Since each
individual subject's regression smooth $f_j(x)$ comes with its own penalization
parameter $\lambda_j$,  subjects' time series of reaction times are effectively
treated as fixed, just as the slopes of by-subject regression lines in the
following linear mixed model are fixed:
\[
y_{ij} = \beta_0 + b_j + \beta_j x_i + \epsilon_{ij} 
\text{ where } \epsilon_{ij} \underset{\text{ind}}{\sim} N(0,\sigma^2) \text{ and } 
  b_j \underset{\text{ind}}{\sim} N(0,\sigma_b^2).
\]
A linear mixed model with random intercepts and random slopes,
\[
y_{ij} = \beta_0 + \beta_1 x_i + b_{0j} + b_{1j} + \epsilon_{ij} 
\text{ where } \epsilon_{ij} \underset{\text{ind}}{\sim} N(0,\sigma^2) \text{ and } 
\bm{b}_j  = \left[ \begin{array}{c} b_{0j} \\ b_{1j} \end{array} \right] \sim N(\bm{0}, \bm{\Psi}), 
\]
\[
\text{ with } \bm{\Psi} = \left[ \begin{array}{cc} \sigma_0 & \rho\sigma_0\sigma_1 \\ \rho\sigma_0\sigma_1 & \sigma_1 \end{array}\right],
\]
has as non-linear equivalent a {\sc gamm} with a factor smooth interaction
(henceforth abbreviated to {\em factor smooth}).  A factor smooth implements
two measures to ensure that effects are proper random effects.  First, a single
smoothing parameter $\lambda$ is used for each of the subject-specific smooths
for trial, forcing penalization to shrink the parameters of the basis functions
in the same way for all subjects. Second, penalization is allowed to affect the
second (completely smooth) basis function that under standard penalization for
wiggliness would not have been effected. This is achieved through additional
penalization of the penalty null space.  In the situation that there is true
wiggliness, a factor smooth will capture this.  When there is no wiggliness,  a
factor smooth will return random intercepts.  

The {\tt edf} values listed in the tables in the appendix for F-tests on the
smooths are based on the sum of the {\tt edf}s of their basis functions, from
which the {\tt edf} for the intercept (equal to 1) has been subtracted, as the
intercept is evaluated separately as a parametric term in the model.  The
p-values listed in these tables for the smooth terms are based on specific $F$
ratios that are discussed in detail in \citet{Wood:2013a} for regression
splines and in \citet{Wood:2013b} for random effects.  The tests for the
regression splines are conditional on the estimates for the smoothing
parameters for other splines in the model.  The test for random effects treats
the variance components that are not tested as fixed at their estimates.  This
assumption makes it possible to test for a zero effect in a computationally
efficient way.  \citet{Wood:2013b} points out that this test is likely to be
less reliable under three circumstances: when variance parameters are not well
estimated, when the assumption that the posterior modes follow a normal
distribution is violated, and when covariates in small samples are highly
correlated. Especially for logistic models with small sample sizes and
correlated covariates, caution is required when p-values are around the
threshold for accepting or rejecting a random effect as significant.

\subsection{Multivariate splines}

Thus far, we have considered univariate smooths $f(x)$, but multivariate
regression splines $f(x_1, x_2, \ldots)$ are also available.  By way of
example, we consider lexical decision latencies for 15,021 Vietnamese compound
words in a single-subject experiment reported in \citet{Pham:Baayen:2015}.
(The data for the analyses reported here and in subsequent sections are, unless
specified otherwise, available in the {\tt RePsychLing} package for {\tt R} at
\url{https://github.com/dmbates/RePsychLing}.)

\begin{figure}[h]
  \centering
  \includegraphics[width=0.9\textwidth]{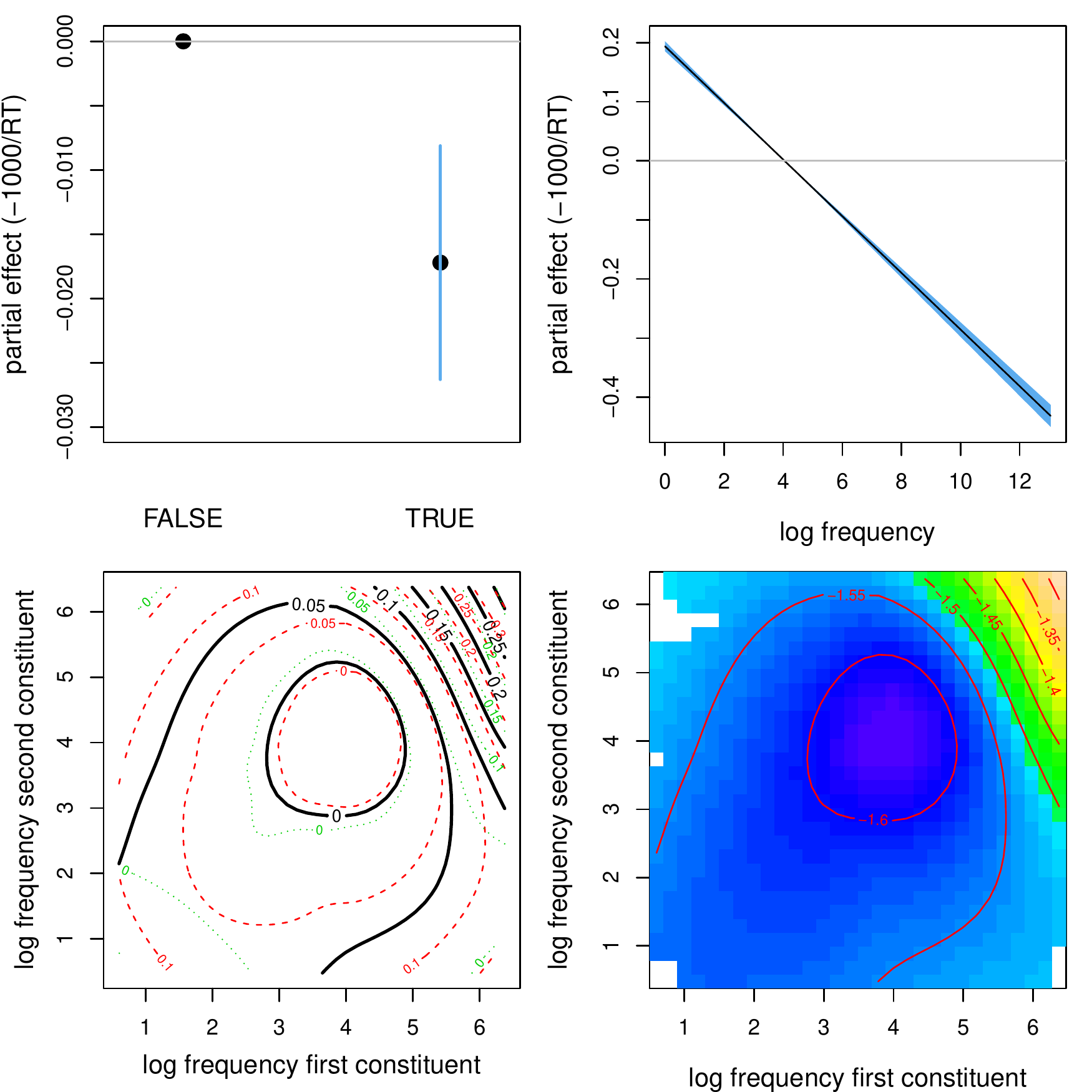}

  \caption{A simplified model for Vietnamese lexical decision latencies to
  compound nouns, with an effect of tone (upper left), word frequency (upper
  right), and the left and right constituent frequencies (bottom panels).  The
  lower left panel shows contour lines with 1SE confidence intervals, the lower
  right panel presents the corresponding contour plot.  Lighter and warmer
  colors denote longer response latencies.}

  \label{fig:vietnamese}
\end{figure}

\clearpage

\begin{figure}
\centering
\includegraphics[width=1.0\textwidth]{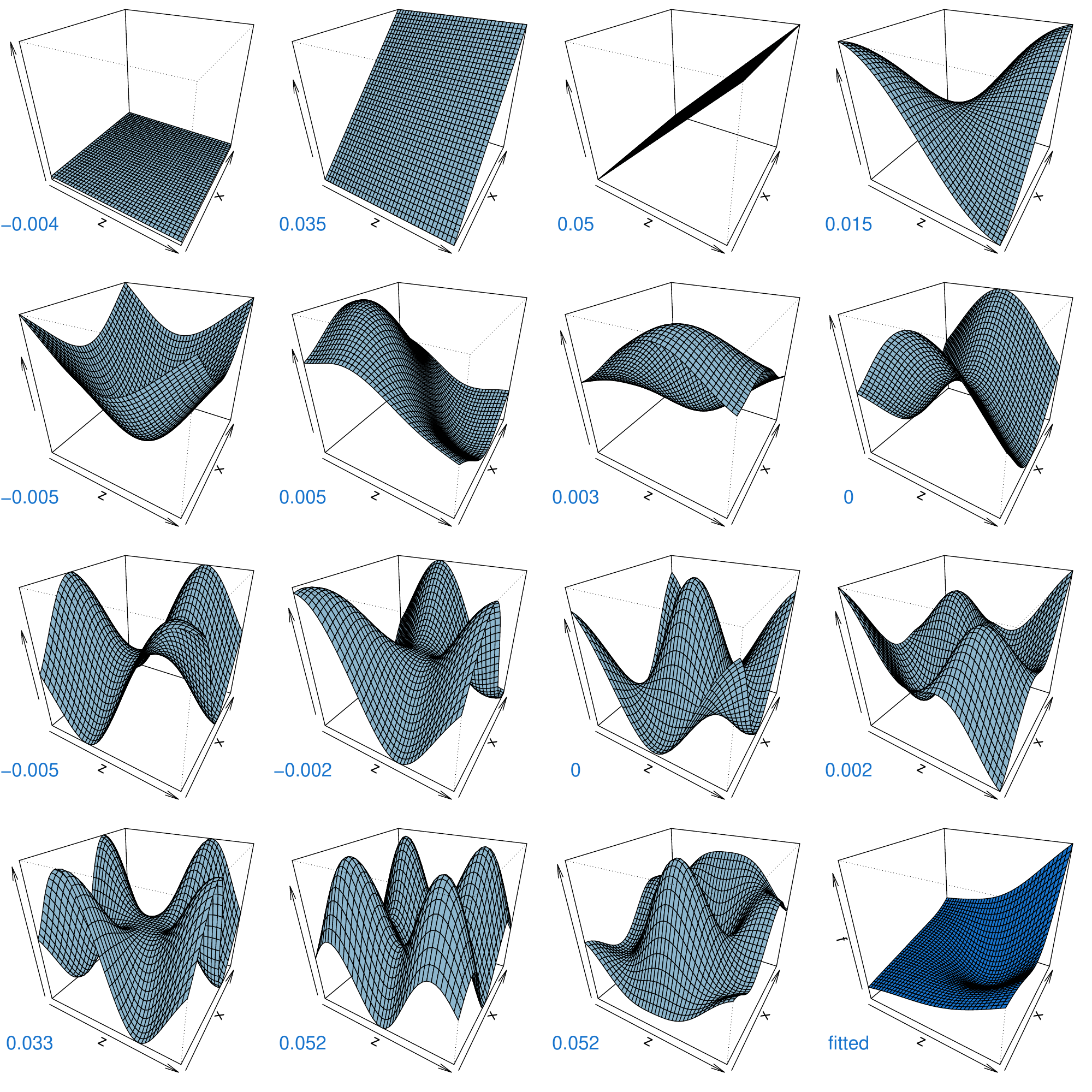}

\caption{Basis functions and their weights for a thin plate regression spline
for the interaction of the constituent frequencies in the {\sc gam} fitted to
the Vietnamese lexical decision data.}

\label{fig:bivarTPRS}
\end{figure}

\clearpage

Figure~\ref{fig:vietnamese} presents a (simplified) model for these visual
lexical decision latencies with four predictors.  First, we include a factor
specifying whether the tone realized on the first constituent of a Vietnamese
compound is the most common (mid-level) tone, or another of six tones. We use
the notation $t(i)$ to denote the level of $\alpha$ for observation $i$, with
as possible values {\sc true} and {\sc false}. The effect of tone
($\alpha_{t(i)}$) is coded with treatment coding.  Second, we included the
frequency of the compound word ($x_{1}$).  Additional predictors were the
frequencies of the first and second constituents ($x_{2}$ and $x_{3}$).  We
used a univariate thin plate regression spline for compound frequency, and we
used a tensor product smooth for the interaction of the two
constituent frequencies:
\[
y_{i} = \beta + \alpha_{t(i)} + f_1(x_{1i}) + f_2(x_{2i}, x_{3i}) +  \epsilon_{i},
\text{ where } \epsilon_{i} \underset{\text{ind}}{\sim} N(0,\sigma^2). 
\]
Before discussing the details of bivariate splines, first consider
Figure~\ref{fig:vietnamese}, which presents the {\em partial effects} of the
predictors, i.e., the contributions of the individual terms in the model.  The
upper left panel visualizes the effect of tone. The reference level (some other
tone than the mid tone) is at zero.  Words with a mid tone on the first
constituent are responded to -0.0172 units faster on the -1000/RT scale.  The
group means for tone (with the other predictors held at their most typical
values) are obtained by adding the intercept, resulting in the estimates
$-1.601$ and $-1.601-0.017 = -1.618$.

Frequency was entered into the model with a thin plate regression spline, but
its effect is linear, and it is this linear effect that is returned by the
regression spline.  The confidence intervals for the line have width zero where
they intersect with the horizontal line crossing 0 on the y-axis.  This is
because for a {\sc gam} to be identifiable, all uncertainty about the intercept
is already quantified through the standard deviation for the intercept.  Since
the upper right panel shows the partial effect of word frequency, the
regression line has to be shifted by the value of the intercept ($-1.601$) to
position it at its appropriate vertical position familiar from standard graphs
of regression lines.  After this shift, the word frequency $f_0$ for which the
regression line crosses the horizontal axis now has the intercept as new $y$
value. But as all uncertainty about the intercept is bundled into the 
standard deviation estimated for the intercept, there is no uncertainty left
about the contribution of $f_0$ to the model's prediction.  As a consequence,
the confidence interval for the partial effect of frequency is zero at $f_0$.

The bottom panels of Figure~\ref{fig:vietnamese} visualize the interaction of
the two constituent frequencies.  Shortest responses are found for intermediate
values, the longest response times occur when both frequencies are high.  The
lower left panel shows contour lines with 1SE confidence intervals, red dashed
lines represent the lower interval, and green dotted lines the higher interval.
The contour plot in the lower right facilitates interpretation with color
coding.  Deeper shades of blue indicate shorter reaction times.

This interaction can be modeled with the help of a bivariate thin plate
regression spline or with a tensor product smooth.  A thin plate regression
smooth $y = f(x, z)$ models a wiggly surface as a weighted sum of simpler
surfaces, as illustrated in Figure~\ref{fig:bivarTPRS}.  There are three
completely smooth surfaces, a horizontal flat plane and two tilted planes. The
remaining surfaces (from left to right and top to bottom) are increasingly
wiggly.  The weighted sum of these surfaces results in the surface in the lower
right, the predicted surface for the interaction of the two constituent
frequencies ($x$: first constituent, $z$: second constituent).  Just as for
univariate thin plate regression splines, penalization ensures a proper balance
between oversmoothing and undersmoothing.

\begin{figure}
\centering
\includegraphics[width=0.5\textwidth]{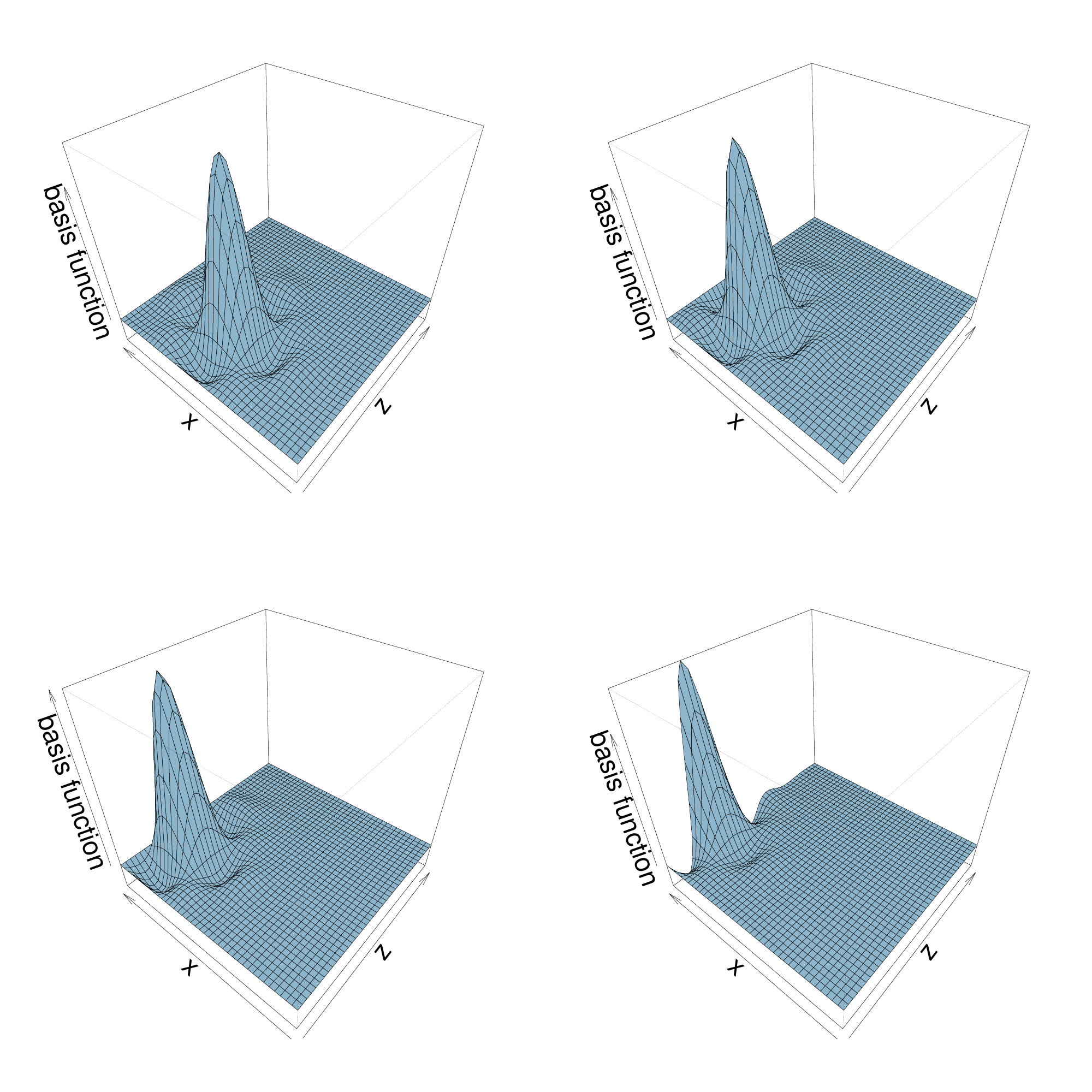}\includegraphics[width=0.5\textwidth]{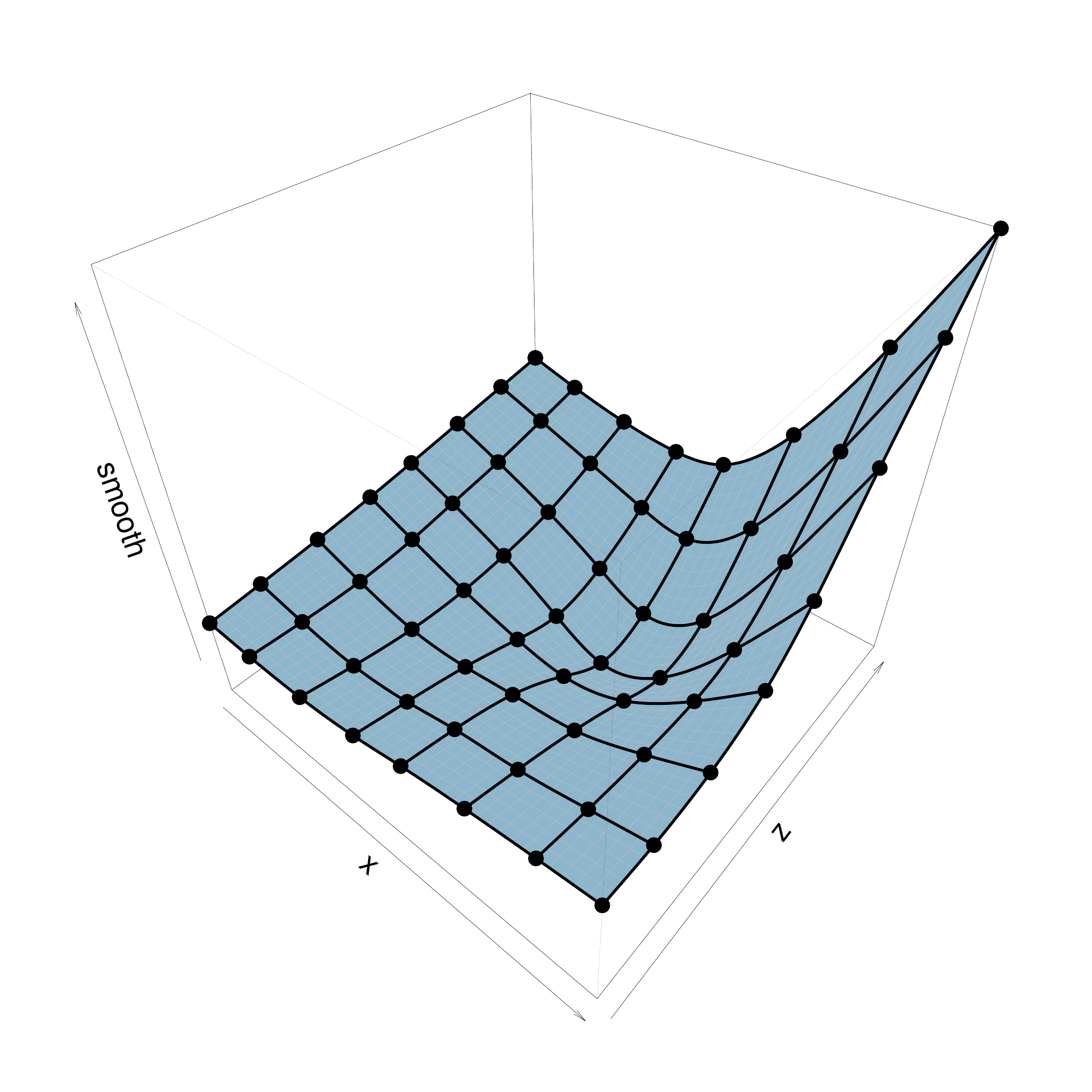}

\caption{
  Examples of the basis functions used in bivariate tensor product smooths are
  illustrated to the left.  The right illustrates the predicted wiggly surface
  for the partial effect of the constituent frequencies in the {\sc gam} fitted
  to the Vietnamese reaction times.  Each dot represents the top of a weighted
  basis function.
}

\label{fig:bivarTENSOR}
\end{figure}

Multivariate thin plate regression splines are appropriate for isometric
predictors, i.e., predictors that are measured on the same scale, such as
longitude and latitude, or first and second constituent frequency.  When
predictors are on different scales, thin plate regression splines cannot be
used.  For interactions of non-isometric predictors, tensor product smooths are
available.  A bivariate tensor smooth makes use of basis functions that
are the three-dimensional counterpart of the two-dimensional basis functions
shown in Figure~\ref{fig:cr} for the univariate case.  These basis functions
are illustrated in the left-hand side of Figure~\ref{fig:bivarTENSOR}.  When 8
basis functions are selected for both predictor dimensions, a total of $8
\times 8 = 64$ basis functions is set up.  Each basis
function is weighted, resulting in predicted values $y$ represented in
Figure~\ref{fig:bivarTENSOR} by $64$ black dots, each representing the maximum
of the basis function for the corresponding knot. Penalization requires two
smoothing parameters, one for each dimension, and is implemented such that the
black curves parallel to the X-axis, and those parallel to the Z-axis, are
properly constrained (see Wood, 2006, chapter 4, for further details).  Tensor
product smooths applied to isometric predictors tend to produce similar results
as thin plate regression splines, but for isometric predictors, thin plate
regression splines tend to offer more precision.  For the Vietnamese compounds,
the two smooths predict regression surfaces that are nearly indistinguishable.

\subsection{Interactions with factorial predictors}

It is often the case that a covariate has a functional form that differs for
the individual levels of a factor.  Different wiggly curves or wiggly
(hyper)surfaces can be fitted to each factor level, as in the model
\[
y_{i} = \beta + \alpha_{t(i)} + f_1(x_{1i}, \text{by}=t(i)) + f_2(x_{2i}, x_{3i}, \text{by}=t(i)) +  \epsilon_{i},
\text{ where } \epsilon_{i} \underset{\text{ind}}{\sim} N(0,\sigma^2), 
\]
where $f_i(x_1, x_2, \ldots, \text{by}=t(i))$ denotes the smooth for the
interaction of $x_1, x_2, \ldots$ by $\alpha$.  For models with these kind of
interactions, the main effect of the factor ($\alpha_{t(i)}$) is an important
component of the model, as it has the crucial function of properly calibrating
the different curves, surfaces (or hypersurfaces) with respect to the
intercept.  {\sc Gam}s can also be set up to estimate the difference between
curves or (hyper)surfaces.

The analyses in this study were carried out with the help of the {\tt mgcv}
package, version 1.8-12 \citep{Wood:2006,Wood:2011} and the {\tt itsadug}
package \citep{itsadug:2016} for {\tt R} \citep[version
3.2.2,][]{Rcite}.\footnote{Except for the {\tt baldey} dataset, which is available 
on-line as documented below, all data sets discussed are available in the 
{\tt RePsychLing} package for {\tt R}. 
Detailed R code for the analyses reported is available in this package in the
folder {\tt inst} as caveOfShadows.pdf, and at
\url{http://www.sfs.uni-tuebingen.de/~hbaayen/publications/supplementCave.pdf}.
} 
These analyses are all exploratory, in that a sequence of increasingly
complex models was constructed and only those predictors and interactions were
maintained that received substantial support for improving the model fit.

This completes the introduction to the generalized additive (mixed) model.  We
now return to the central topic of this study, and turn to the first dataset
illustrating that experimental data can be infected by the human factor in a
non-trivial way.

\section{The human factor in three experiments}

\subsection{The {\tt KKL} dataset}

\begin{figure}
  \centering
  \includegraphics[width=0.8\textwidth]{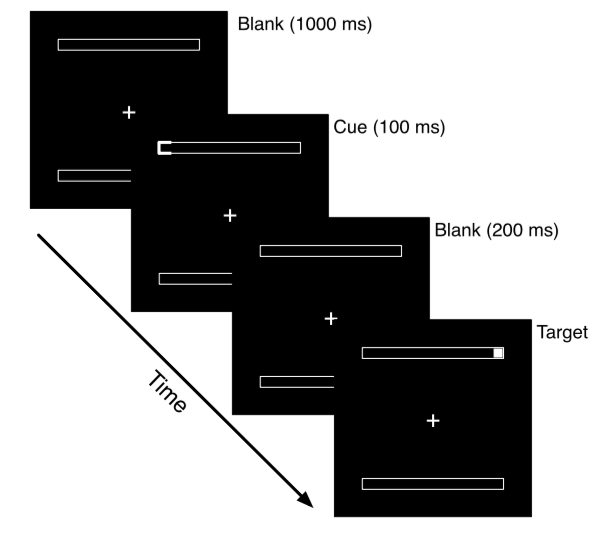}

  \caption{Sequence of events in visual-spatial attention experiment with an
  invalid cue on the same object. Screen 2: Left end of top rectangle is cued.
  Screen 3: SOA of 200 ms, Screen 4: Large target to be detected at right end
  of top rectangle (from Kliegl et al., 2011). In the new experiment rectangles
  were also presented in diagonal orientations; a different group of subjects
  was tested with small targets.}

  \label{fig:KKLprocedure}

\end{figure}

The experiment reported by \citet{Kliegl:Kuschela:Laubrock:2015}, a follow
up to \citet{Kliegl:Wei:Dambacher:Yan:Zhou:2011}, showed that validly cued
targets on a monitor are detected faster than invalidly cued ones, i.e.,
spatial cueing effect \citep{Posner:1980} and that targets presented at the
opposite end of a rectangle at which the cue had occurred were detected faster
than targets presented at a different rectangle but with the same physical
distance, an object-based effect \citep{Egly:Driver:Rafal:1994}.  The sequence
of an experimental trial is shown in Figure~\ref{fig:KKLprocedure}.  Different
from earlier research, the two rectangles were not only presented in cardinal
orientation (i.e., in horizontal or vertical orientation), but also diagonally
(45 degrees left or 45 degrees right).  This manipulation afforded a follow up
of a hypothesis that attention can be shifted faster diagonally across the
screen than vertically or horizontally across the screen
\citep{Kliegl:Wei:Dambacher:Yan:Zhou:2011,Zhou:Chu:Li:Zhan:2006}.  Finally,
data are from two groups of subjects, one group had to detect small targets and
the other large targets.  For an interpretation of fixed effects relating to
the speed of visual attention shifts under these experimental conditions we
refer to \citet{Kliegl:Kuschela:Laubrock:2015}.

Eighty-six subjects participated in this experiment. There were 800 trials
requiring detection of a small or large rectangle and 40 catch trials. The
experiment is based on a size (2) $\times$ cue-target relation (4) $\times$
orientation (2) design. Targets were small or large; rectangles were displayed
either in cardinal or diagonal orientation, and cue-target relation was valid
(70\% of all trials) or invalid in three different ways (10\% of trials in each
of the invalid conditions), corresponding to targets presented (a) on the same
rectangle as the cue, but at the other end, (b) at the same physical distance
as in (a), but on the other rectangle, or (c) at the other end of the other
rectangle. Size of target was varied between subjects, the other two factors
within subjects. The three contrasts for cue-target relation test differences
in means between neighboring levels: spatial effect, object effect, and
gravitation effect \citep{Kliegl:Wei:Dambacher:Yan:Zhou:2011}. Orientation and
size factors are also included as numeric contrasts in such a way that the
fixed effects estimate the difference between factor levels. With this
specification the intercept estimates the grand mean of the 16 ($= 2 \times 4
\times 2$) experimental conditions. The data are available as {\tt KKL} in the
{\tt RePsychLing} package.  The dependent variable is the log of reaction time
for correct trials completed within a 750 ms deadline. The total number of
responses was 53765.

\citet{Bates:Kliegl:Vasishth:Baayen:2015} determined a parsimonious mixed
model for these data, dealing with issues of overparameterization. We refitted
this model using in addition a quadratic polynomial, which allowed us to
include a well-supported nonlinear effect for stimulus onset asynchrony, which
was varied randomly in an interval ranging from 300 to 500 ms in this
experiment, but the effect of which had not been included in the initial {\sc
lmm} report of Bates et al. (2015).

\begin{figure}[ht]
\centering
\includegraphics[width=1.0\textwidth]{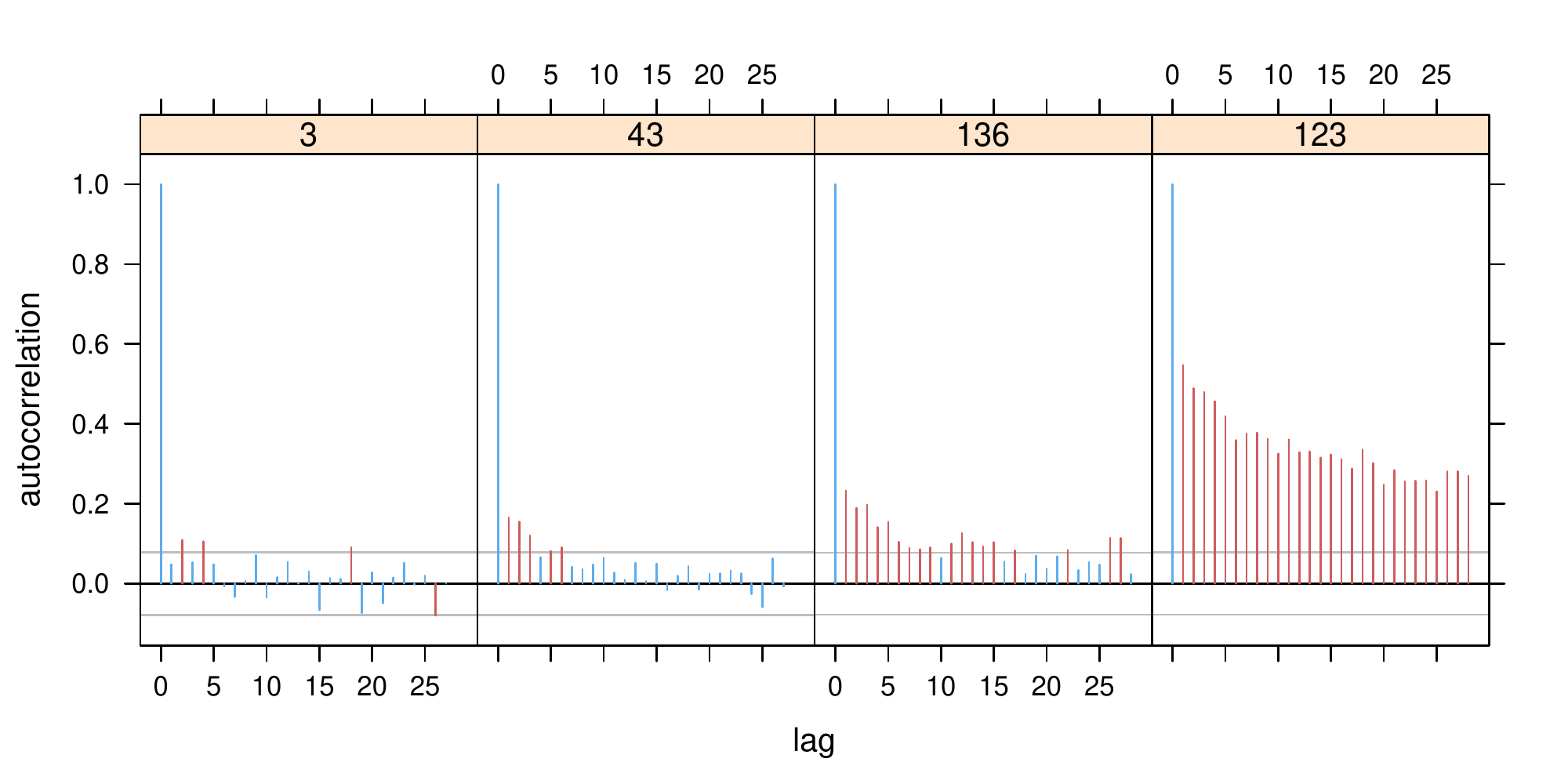}
\includegraphics[width=1.0\textwidth]{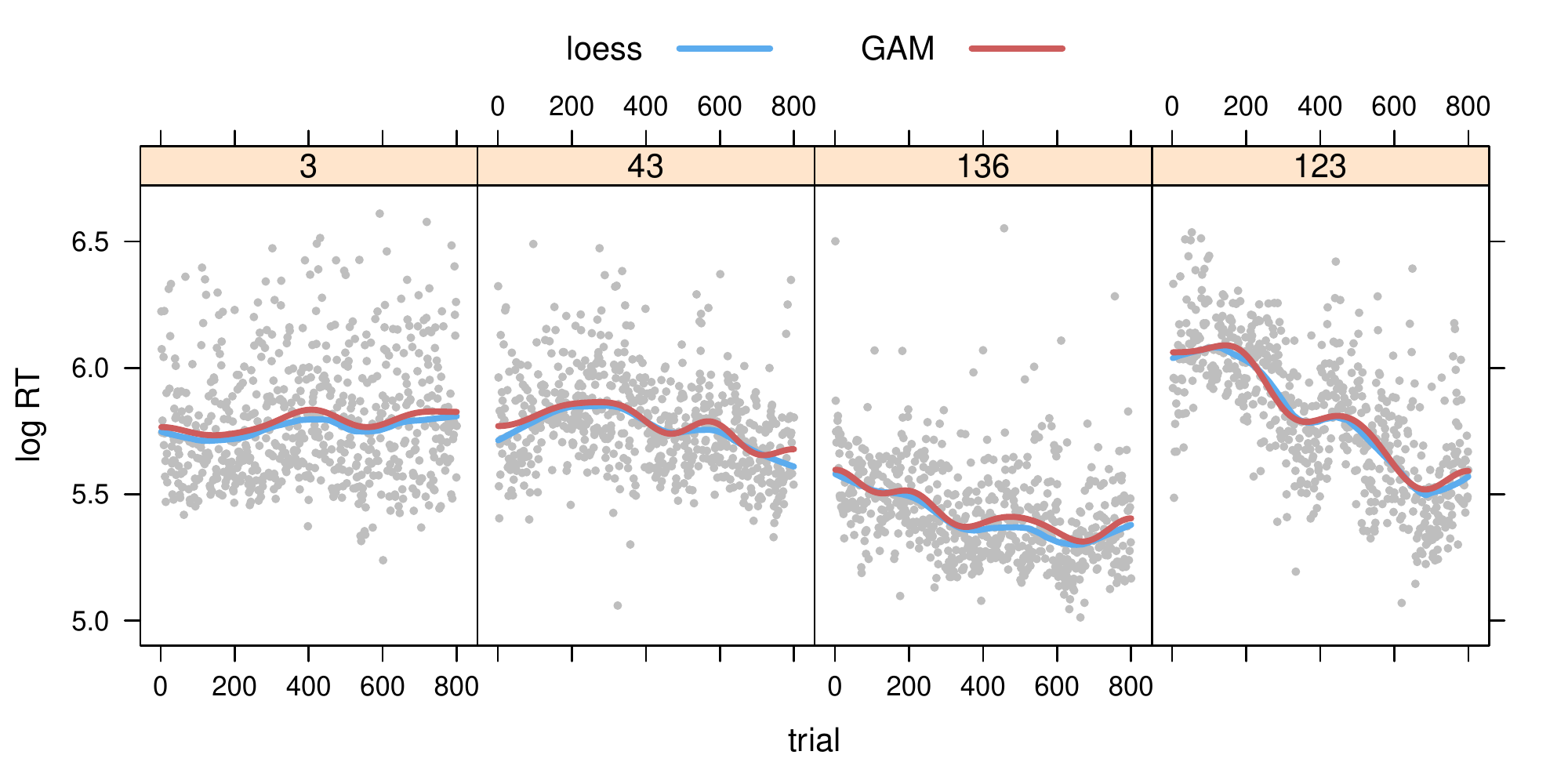}

\caption{Autocorrelation functions for the residuals of a linear mixed model
for four subjects in the {\tt KKL} dataset (upper panels), and the
corresponding plots (lower panels) graphing log reaction time (RT)  against
trial, with a loess smoother (span = 0.2, in blue) and a {\sc gam} factor
smooth (red).}

\label{fig:smoothers}
\end{figure}

Model criticism is an important but all too often neglected part of data
analysis.  Inspection of the residuals of the reference model reveals that
although the residuals approximately follow a normal distribution, and although
they are identically distributed, they are not independent.  

The reaction times of a given subject constitute a time series, with
experimental trial as unit of time.  These trials can be ordered from the
initial trial in the experimental list of trials, to the final trial in that
list.  In what follows, we refer to the time series of trials with the
covariate {\tt Trial} $= 1, 2, \ldots, k$.  For the present experiment, we
have 86 such time series, one for each of the 86 subjects.  When we consider
the residuals of  the reference model, ordered by these time-series, we
observe autocorrelative structure.

The strength of the autocorrelations in these by-subject time series varied
from subject to subject.  For four exemplary subjects in the top panels of
Figure~\ref{fig:smoothers}, the autocorrelation function is shown for the
residuals of a linear mixed model fitted to the {\tt KKL} dataset.  The
autocorrelations for the subject in the top left panel are quite mild, and
unlikely to adversely affect model statistics.  For the second subject, we find
evidence for autocorrelations up to at least lag 3.  Autocorrelations increase
for the third subject, and are still present at a lag of 15 trials.  The
subject in the rightmost panel show strong autocorrelations, indicating that a
response time at trial $t$ is remarkably well correlated with the response at
time $t-L$, for lags $L$ as large as 25.

Given that the residuals of the reference model (refitted with a {\sc gamm})
are not independent, it is unclear how reliable the estimates of model
parameters and the assessments of the uncertainty about these estimates
actually are.  For this particular data set, strong autocorrelations such as
for the last subject are exceptional, and hence it is likely that conclusions
based on this model will be somewhat accurate.  Nevertheless, a statistical
model that is formally deficient is unsatisfactory, especially as there must be
hidden temporal processes unfolding in this experiment that are not transparent
to the analyst.  Since the {\tt KKL} data are clearly not sterile, a more
fertile approach is to bring such hidden processes out in the open, and
incorporate them into the statistical model.  

Why are these autocorrelations present?  In order to address this question,
consider the plots in the lower set of panels of Figure~\ref{fig:smoothers}.
These panels present scatterplots of the data points for the four subjects in
the corresponding top panels, to which two smoothers have been added, a {\sc
loess} locally weighted scatterplot smoother \citep{Cleveland:79} in blue
and a smoother obtained with a generalized additive model in red.  For subject
3, whose time series of responses hardly shows any autocorrelation, we observe
smooths that are close to horizontal lines.  As the experiment proceeds, there
are only very small changes in average response time.  When we move further to
the right in the array of panels, temporal patterns begin to emerge.  As the
experiment proceeded, subjects responded more quickly. Furthermore, it appears
that there may be undulations in response speed.  These oscillating changes in
amplitude, if real, may reflect slow changes in subjects' attention or
concentration over the course of the experiment.  By contrast, the general
downward trend present for subjects 43, 136, and especially 123 may point to
familiarization with and gradual optimization of response behavior for the
task.

\begin{figure}
\centering
\includegraphics[width=0.7\textwidth]{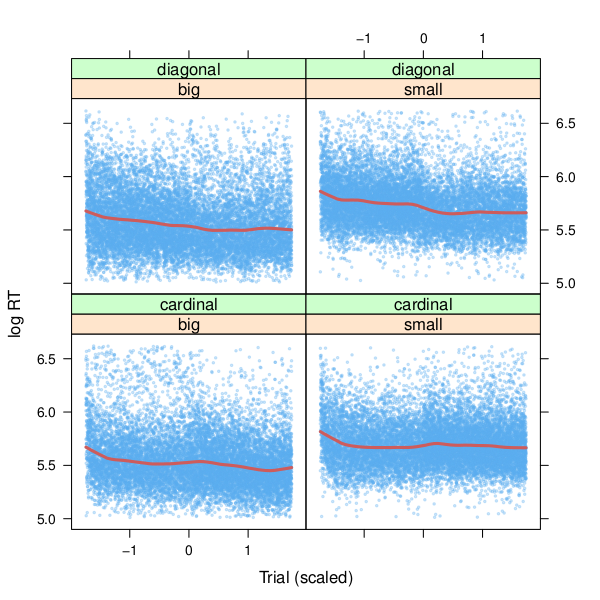}
\caption{Loess smooths (span = 0.2) for the three-way interaction of {\tt Trial} by {\tt Orientation}
by {\tt Size}.}
\label{fig:sizeCardinal}
\end{figure}

The presence of a potential learning effect raises the question of whether
learning proceeded in the same way across the different experimental
conditions.  Graphical exploration suggests that the rate at which subjects
respond faster over time indeed varies, specifically so across the levels of
{\tt size} and {\tt orientation}, as shown in Figure~\ref{fig:sizeCardinal}.  
In the upper panels ({\tt diagonal} orientation), reaction times decrease
over the first half of the experiment and then level off, with a somewhat
greater increase for {\tt small} size.  For {\tt cardinal} orientation, 
reaction times decrease more quickly early on in the experiment, level
off near the middle of the experiment, and then continue their descent
for the condition with size {\tt big}.

Within the context of the linear mixed model, the observed effects of trial can
be taken into account by incorporating by-subject random slopes for {\tt
Trial}, and by allowing {\tt Trial} to interact with {\tt Size} and {\tt
Orientation}.  As shown in Table~\ref{tab:KKLaicsLMER}, these extensions of our
reference model are solidly supported by model comparisons using likelihood
ratio tests.  A summary of the final linear mixed model can be found in
Table~\ref{tab:KKLlmer3} in the appendix.

\begin{table}[ht]
\centering
\begin{tabular}{lrrrrrrr} \hline
                            & Df &       AIC &    logLik & deviance  & Chisq  &Df & Pr($>$Chisq) \\ \hline
reference model             & 25 & -25087.68 &  12568.84 & -25137.68 &        &   &              \\ 
add Trial L * (sze+orn)     & 28 & -25884.64 &  12970.32 & -25940.64 & 802.96 & 3 & $ < 0.0001$ \\ 
add Trial Q * (sze+orn)     & 31 & -26174.41 &  13118.20 & -26236.41 & 295.77 & 3 & $ < 0.0001$ \\ 
add random slopes Trial     & 33 & -26988.91 &  13527.45 & -27054.91 & 818.50 & 2 & $ < 0.0001$ \\ \hline
\end{tabular}
\caption{Model comparisons for linear mixed models fitted to the {\tt KKL} dataset. L: linear
term of a quadratic polynomial; Q: quadratic term of this polynomial; sze: Size; orn: Orientation.}
\label{tab:KKLaicsLMER}
\end{table}

\begin{figure}
\centering
\includegraphics[width=0.7\textwidth]{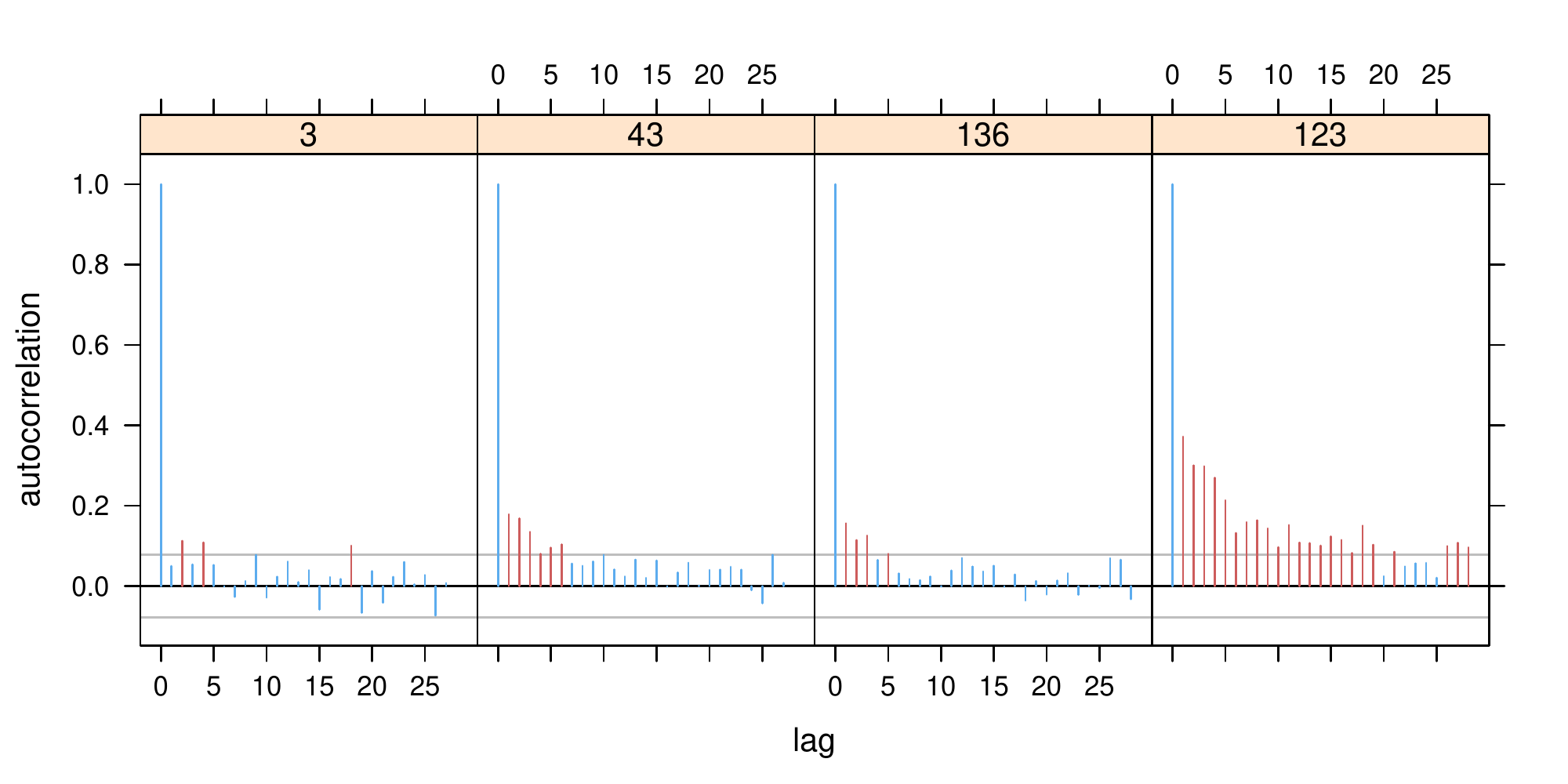}
\caption{Autocorrelations in the residuals of the extended linear mixed model fitted to the {\tt KKL} data.}
\label{fig:KKLacfLMER}
\end{figure}

Figure~\ref{fig:KKLacfLMER} presents the autocorrelation functions for the
residuals of this final, comprehensive, linear mixed model.  Comparison with
the top panels of Figure~\ref{fig:smoothers} shows that for subjects 136 and
123, the autocorrelation in the residuals has been reduced substantially,
thanks to bringing the effects of {\tt Trial} into the model.  Nevertheless,
some autocorrelation remains present.

For further reduction of autocorrelations, it is necessary to relax the
assumption that the subject-specific effects of {\tt Trial}, currently modeled
by means of by-subject random intercepts and random slopes, are strictly
linear.  The smooths presented in the bottom panels of
Figure~\ref{fig:smoothers} suggest that undulations may ride on top of the
linear trends.  What we need, then, is a way of relaxing the linearity
assumption for the by-subject random effects of {\tt Trial}.  The factor smooth
interaction of the generalized additive mixed model provides the required
nonlinear counterpart to the combination of random slopes and random
intercepts.  A factor smooth for {\tt Trial} by subject sets up a separate
smooth for each level of the factor {\tt Subject}.  When we add the constraint
that each smooth should have the same smoothing parameter, and penalize the
smooths for wiggliness, thereby shrinking them towards zero, we obtain `wiggly
random effects'.

The red smooths in Figure~\ref{fig:smoothers} are such factor smooths.  They are
very similar to the loess smooths, but are slightly more sensitive to the
undulations in the data.  Althought it might seem there is a risk that the
factor smooths are modeling noise rather than signal, this is unlikely as the
factor smooths are evaluated within the general framework of the generalized
additive mixed model, and hence it is possible to assess whether they contribute
significantly to the model fit.  By way of example, across 100 random
permutations of {\tt Trial} for the four subjects of Figure~\ref{fig:smoothers},
a significant factor smooth was obtained in 3 instances for $\alpha=0.05$ and
for zero cases for $\alpha = 0.01$, indicative of nominal Type~I error rates.
This example illustrates informally that factor smooths are unlikely to find
and impose nonlinear structure when there is none.   

Importantly, we think the undulating random effects captured by factor smooths
represent the ebb and flow of attention.  They emerge not only in the present
data set, but have been observed for visual lexical decision
\citep{Mulder:Dijkstra:Schreuder:Baayen:2014} as well as for word naming and
for {\sc eeg} data \citep{Baayen:VanRij:DeCat:Wood:2015}.  If this
interpretation is correct, penalized factor smooths are the appropriate
statistical tool to use.  We explicitly do not want to model these fluctuations
in attention as fixed effects, because there is no reason to believe that if
the experiment were replicated, a given subject would show exactly the same
pattern.  It is more realistic to expect that changes in attention will again
be present, with roughly the same magnitude, but with ups and downs occurring
at different points in time.  In other words, we are dealing here with
temporally structured noise, and the penalized factor smooths make it possible
to bring such `wiggly random effects' into the statistical model.

\begin{table}[ht]
\centering
{\footnotesize
\begin{tabular}{lrrrlrrr} \hline
  &       {\sc aic} &     f{\sc reml} & Df & comparison with & Chisq     & Df difference  & $\Pr(>\text{Chisq})$ \\ \hline
 reference model   & -26009.55 & -12495.77 & 27 &                 &           &                &            \\                  
 linear model      & -28047.29 & -13422.25 & 34 & reference model &   926.5   &    7           & $<$ 0.0001 \\                  
 factor smooths    & -30876.72 & -14500.08 & 31 & linear model    &  1077.8   &   -3           & $\dagger$  \\                  
 smooth trial      & -31040.29 & -14582.64 & 33 & factor smooths  &    82.6   &    2           & $<$ 0.0001 \\ \hline           
\end{tabular}
}

\caption{Model comparison for {\sc gamm}s fitted to the {\tt KKL} dataset.
(The reference and linear models were refitted with {\sc gamm}s to ensure
comparability across all four models.) $\dagger$: the model with factor smooths
has lower f{\sc reml} (\marked{fast REML}, see \texttt{mgcv} documentation) 
and fewer parameters, and thus is simpler and better, than the linear model.}

\label{tab:modCompGAMM}
\end{table}

Table~\ref{tab:modCompGAMM} lists four {\sc gamm}s that we fitted to the {\tt
KKL} data.  The first is the reference model, but refitted with {\sc gam}
software (the {\tt mgcv} package for {\tt R}), rather than with {\sc lmm}
software (the {\tt lme4} package for {\tt R}), with as the only change that a
thin plate regression spline is used for the {\tt SOA} covariate, instead of a
quadratic polynomial.  The second model has the same specification as the final
linear mixed model (summarized in Table~\ref{tab:KKLlmer3} in the appendix),
but refitted with a {\sc gamm}.  (These two models were refitted because both
the estimation algorithms and the way in which degrees of freedom are handled
differ between {\tt lme4} and {\tt mgcv}.)  As expected, the linear model,
which includes effects for {\tt Trial}, outperforms the reference model.   The
third model, which replaces the by-subject random intercepts and slopes by
factor smooths, provides a better fit with fewer effective degrees of freedom.
Addition of the three-way interaction of {\tt Trial} by {\tt Size} and {\tt
Orientation} improves the model further.

\begin{figure}
\centering
\includegraphics[width=0.7\textwidth]{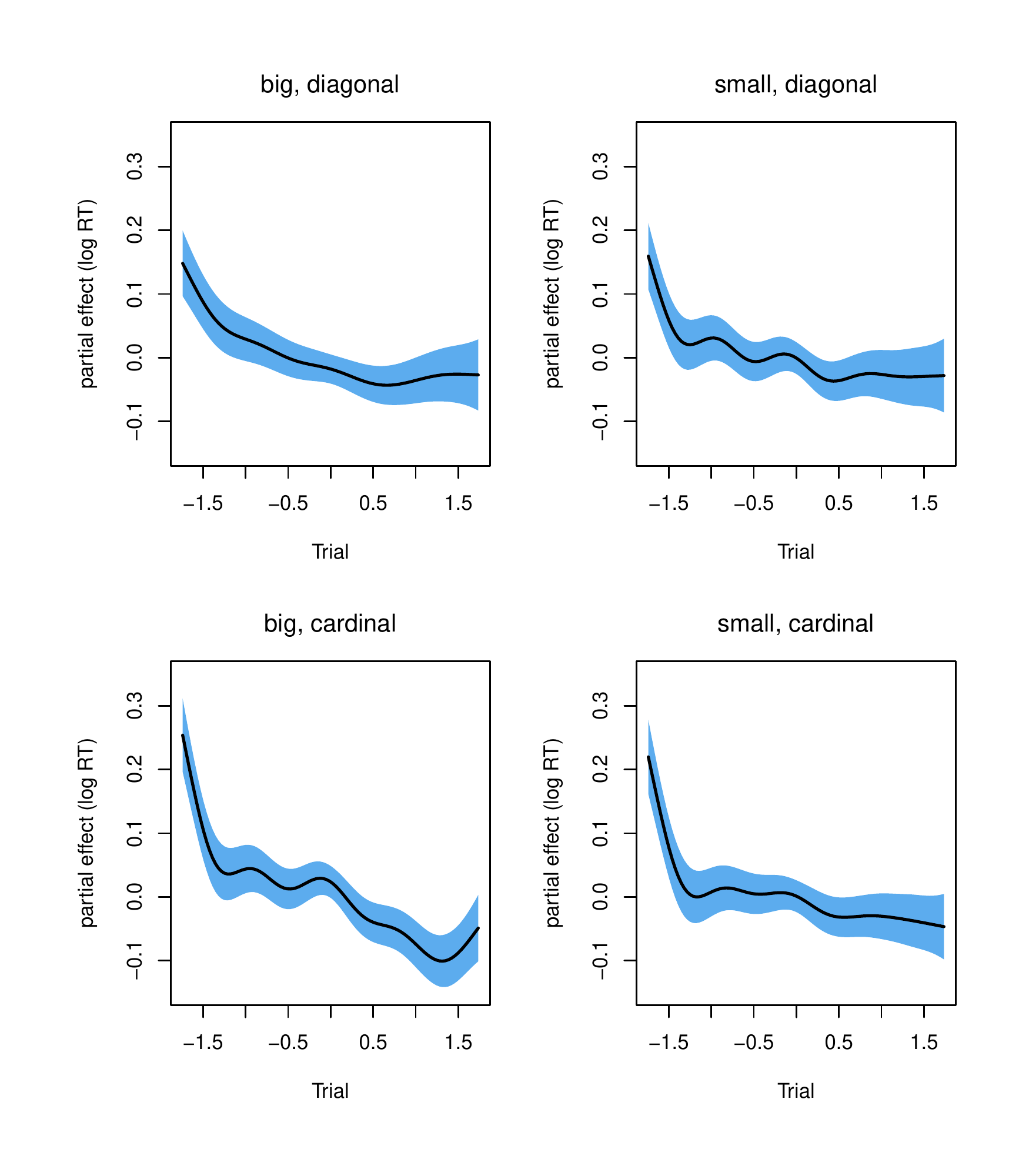}
\caption{The {\tt Trial} by {\tt Orientation} by {\tt Size} interaction 
in the full {\sc gamm} model for the {\tt KKL} data.}
\label{fig:TrialFinGamm}
\end{figure}

Figure~\ref{fig:TrialFinGamm} visualizes this three-way interaction of {\tt
Trial} by {\tt Orientation} by {\tt Size} estimated by the {\sc gamm} (cf.
Figure~\ref{fig:sizeCardinal} for the corresponding loess smooths).  The effect
of learning is larger for the cardinal presentation (upper panels) than for the
diagonal presentation (bottom panels).  For both large (left panels) and small
(right panels) stimuli, we observe rapid initial learning, which levels off
more for small than for large stimuli.  Big stimuli with diagonal presentation
elicited the most smooth accommodation pattern, with response times gradually
becoming shorter.

\begin{figure}
\centering
\includegraphics[width=0.7\textwidth]{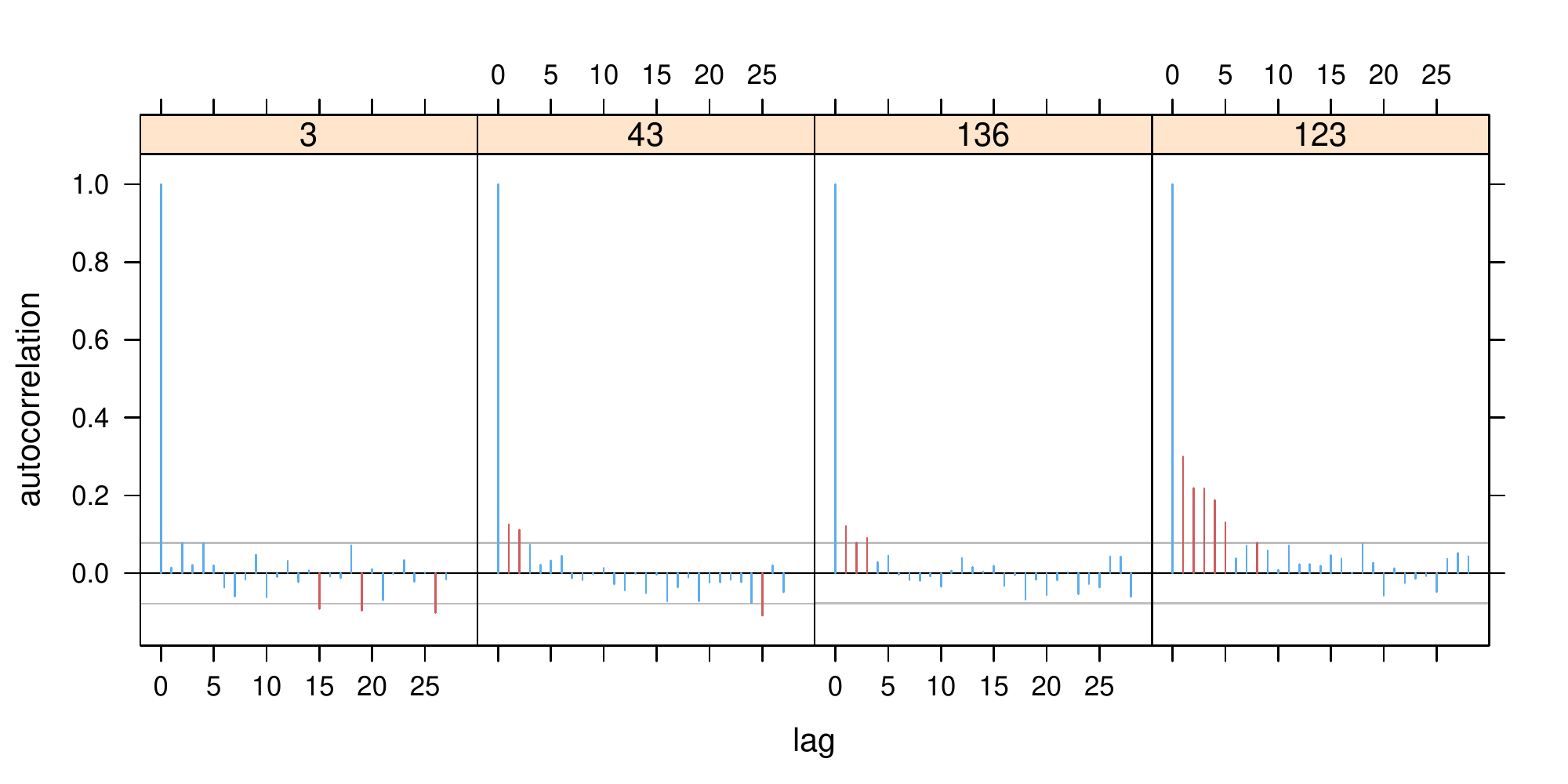}
\caption{Autocorrelation functions for four subjects in the full {\sc gamm} 
fitted to the {\tt KKL} data.}
\label{fig:KKLacfGAMM}
\end{figure}

Figure~\ref{fig:KKLacfGAMM} clarifies that the full {\sc gamm} succeeded in
further reducing the autocorrelations in the residuals.   This reduction is due
almost exclusively to the use of factor smooths, with only tiny amelioration by
adding in the three-way interaction with {\tt Trial}.  This result is important
for two reasons.  First, it is unlikely that the removal of autocorrelation in
the residuals could be accomplished by factor smooths fitting noise rather than
signal.  Likewise, it is unlikely that the huge reduction in {\sc aic} when
going from the linear mixed model to the {\sc gamm} (2087, see
Table~\ref{tab:modCompGAMM}) could be accomplished by just fitting noise.
Second, the presence of slowly undulating processes in experimental data is a
phenomenon that is itself of theoretical interest, and invites interpretation,
clarification, and replication.

The remaining autocorrelations that are visible in Figure~\ref{fig:KKLacfGAMM}
are unlikely to be harmful, but to play safe, one might consider removing
subject 123 from the dataset and refitting the model.  Alternatively, these
last remaining autocorrelations might be due to a simple {\sc ar(1)}
autocorrelative process in the errors, according to which the current error
$e_t$ in the timeseries at time $t$ is equal to a proportion $\rho$ of the
preceding error $e_{t-1}$ plus Gaussian noise $\epsilon_t$:
\begin{equation}
  e_t = \rho e_{t-1} + \epsilon_t,
  \text{ where } \epsilon_t \underset{\text{ind}}{\sim} N(0,\sigma^2). 
\end{equation}
\citet{Pinheiro:Bates:2000} and \citet{Galecki:Burzykowski:2013} provide
extensive discussion of how autocorrelation processes can be accounted for
within the mixed modeling framework; \citet{wood2015generalized} provides
technical details for {\sc gamm}s.  With a mild proportionality constant $\rho
= 0.15$, autocorrelations are almost completely removed. Below, we will discuss
the use of this parameter in more detail.  {\color{black} Here, we note that
models from which the factor smooths are removed, and for which $\rho$ is
increased, provide substantially worse fits to the data, and fail to remove
substantial autocorrelational structure at longer lags. This shows that the
factor smooths may be essential for bringing under statistical control a
substantial part of autocorrelative structure in experimental data.} 

Addressing the autocorrelation issue for the {\tt KKL} data set does not lead
to major changes in significances and magnitudes of fixed-effect coefficients
and the magnitudes of the (significant) coefficients, as reported in
\citet{Bates:Kliegl:Vasishth:Baayen:2015} and
\citet{Kliegl:Kuschela:Laubrock:2015}.  Nevertheless, the {\sc gamm} offers
enhanced insight into the data, specifically with respect to the effect of {\tt
Orientation}.  In the reference model, the coefficient of this main effect was
estimated at 0.041, with a standard error of 0.010 ($t = 3.9$).  However, in
the full linear mixed effect model, the coefficient is smaller (0.014), comes
with greater uncertainty (standard error 0.09), and is \marked{not significant}
($t = 1.5$).  However, the final {\sc gamm} estimates the coefficient at 0.039,
with a standard error of 0.016 and a $t$ value of 2.491, reporting $p = 0.013$.
Furthermore, the interaction of {\tt Size} by {\tt Orientation}, which is
significant in the {\sc lmm}, is not significant in the {\sc gamm}.  Given the
interaction of {\tt Trial} by {\tt Orientation}, the greater uncertainty about
this main effect in the linear mixed model, and about the interaction with {\tt
Size}, makes sense.   Furthermore, as expected, the variance of the conditional
modes for the by-subject random effects of {\tt Orientation} in the reference
model is larger (by a factor two) than in the final {\sc gamm} ($p < 0.0001$,
$F$-test).  Again, this makes sense, as part of what originally looked like
random noise linked to orientation can now be attributed to a learning effect
over experimental time.

In summary, the {\tt KKL} dataset is not sterile, but infected by the `human
factor'.  The by-subject time series are characterized by autocorrelated
errors.  Unlike particles in physics, or plots of barley in agricultural
experiments, human subjects are intelligent beings whose behavior is not random
over time, but adaptive.  The present reanalysis shows that subjects adapt in
different ways to the novel manipulation of canonical versus diagonal
positioning of visual stimuli, which is a theoretically fertile result.  
This result is not available under one-size-fits-all mechanical model selection
procedures based on the a-priori assumption that the data are sterile.

\subsection{The baldey dataset}

Our second example addresses the analysis of the response latencies elicited
in the auditory lexical decision megastudy of \citet{ernestus2015baldey} (data
available at \url{http://www.mirjamernestus.nl/Ernestus/Baldey/baldey_data.zip}).
Ernestus and Cutler include in their statistical analysis the reaction time
to the preceding trial as a way of controlling for temporal dependencies
in by-subject time series.  As shown by \citet{Baayen:Milin:2010}, the
inclusion of preceding reaction time successfully removes a considerable
amount of autocorrelation in the residuals.  Ernestus and Cutler also
included {\tt Trial} as a main effect, together with by-subject random
slopes for {\tt Trial}.

In what follows, we present an analysis of the reaction times of the {\tt
baldey} data, which shows that the human factor is even stronger in these data
than suggested by the analyses of Ernestus and Cutler.

In our analysis --- which is far removed from a ``comprehensive'' analysis of
this rich data set --- we departed from the analyses of Ernestus and Cutler in
several ways.  First, we analyzed an inverse transform of the reaction times
(-1000/RT) rather than a logarithmic transform, as both graphical inspection
and an analysis following \citet{Box:Cox:1964} indicated the inverse
transformation to better approximate normality.

Second, autocorrelations in the errors should not be brought under control by
including preceding reaction time as a covariate.  A statistical model with
preceding reaction time as predictor is no longer a generating model, in the
sense that it is no longer possible, given the model specification, to simulate
the reaction times.  We therefore explored by-subject factor smooths in
combination with the possibility of an {\sc ar(1)} process in the errors.

Third, we relaxed the assumption that effects would be the same across the two
genders.  There are indications that males and females may be differentially
sensitive to word frequency \citep{Ullman:BL:2002,Balling:Baayen:2008}, 
but a gender by frequency interaction is not always found 
\citep{Balling:Baayen:2012,Tabak:Schreuder:Baayen:2005,Tabak:Schreuder:Baayen:2010}.
As the {\tt baldey} data set combines a perfectly balanced set of subjects (10
males and 10 females) with a large number of items (2780 Dutch words), it
provides a testing ground for differential effects of the two genders
in lexical processing.

Finally, we relaxed linearity assumptions, replacing a strictly linear
mixed model by a generalized additive mixed model.

The random effects structure of the model for the reaction times for words (see
Table~\ref{tab:baldey} in the appendix for a statistical summary) included
random intercepts for word, as well as by-word random intercepts for gender.
Different words enjoy different popularity across the genders 
\citep[see also][] {Baayen:VanRij:DeCat:Wood:2015}, 
and adjusting by-word intercepts for
gender results in a tighter model fit.   With respect to subject, we included
by-subject factor smooths for session.\footnote{
Given the small number of sessions (11) and the large number of observations for
each session (around 4500 for each subject), one could opt for treating session
as a factor, and including by-subject random slopes for this factor. Results
are very similar, with the factor smooths showing slightly more shrinkage.
}
The data for this mega-study were collected over 11 sessions, and once 
by-subject factor smooths for session were included in the model, by-subject
factor smooths for {\tt Trial} became redundant.  For subject, random slopes
for the acoustic duration of the stimulus word were also well supported.

\begin{figure}
\centering
\includegraphics[width=0.9\textwidth]{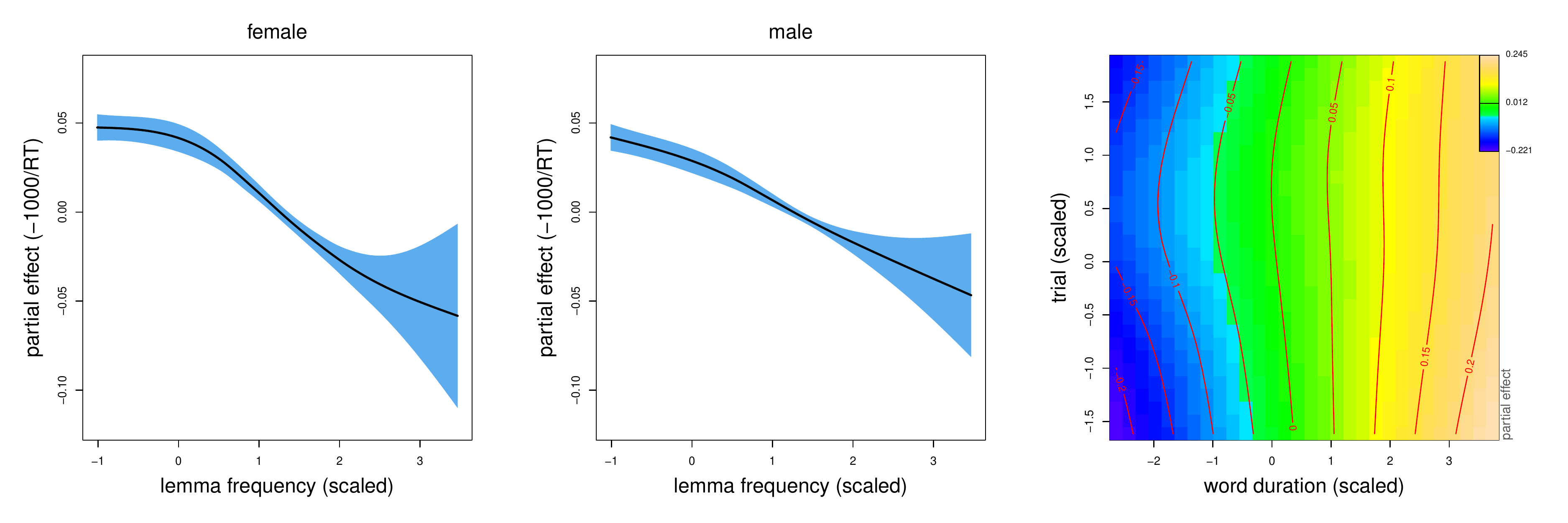}
\caption{Interactions of {\tt Frequency} by {\tt Gender} (left and center panel), and
of {\tt Word Duration} by {\tt Trial} (right panel), in the {\tt baldey} data set.
In the right panel, contour lines are 0.05 units apart (on the -1000/RT scale).
}
\label{fig:baldey}
\end{figure}

Lemma frequency (the summed frequency of a word's inflectional variants)
revealed a non-linear effect that differed between females and males, as shown
in the left and center panels of Figure~\ref{fig:baldey}.  Females show a
somewhat stronger frequency effect, as expected given the somewhat greater
verbal skills of women compared to men \citep{Kimura:2000} and replicating
earlier results \citep{Ullman:BL:2002,Balling:Baayen:2008}.  A novel finding is
that the frequency effect appears linear for men, but shows a curvilinear
pattern for women with little or no effect for very low and very high
frequencies.  Possibly, both the reduced slope and the simpler functional form
of the male curve is tied in with the lesser verbal skills of men.

Furthermore, the effect of the acoustic duration of the auditory stimulus
showed a small but statistically well-supported modulation by {\tt Trial},
visualized in the right panel of Figure~\ref{fig:baldey}.  About two-thirds
through an experimental session, the effect of acoustic duration decreased
somewhat.  This can be seen by noticing the reduced gradient for (scaled) {\tt
Trial} equal to 0.5: the number of contour lines crossed when moving
horizontally across the plot, i.e., for increasing acoustic duration, is
smaller than early on in the experiment.   

As for the {\tt KKL} dataset, inclusion of session and trial in the model did
not absorb all autocorrelation in the residuals.  With $\rho = 0.2$, the
remaining autocorrelations were properly accounted for.

What this analysis shows is that subjects participating in an experiment with
language materials bring with them their own experiences with the language, and
that these specific experiences will lead to differentiated effects that for
the {\tt baldey} data set are partially differentiated by {\tt Gender}.
Furthermore, the effect of acoustic duration varied in the course of the
experiment, providing further evidence for the human subject as a moving target
\citep[see also][]{Ramscar:Hendrix:Love:Baayen:2014,Baayen:Tomaschek:Gahl:Ramscar:2015}.

A strictly linear model provides an inferior fit to the {\tt baldey} data
(f{\sc reml} linear model: \mbox{-13027.88}, f{\sc reml} {\sc gamm}: -14911.48,
approximate test (informal because the models are not strictly nested):
$\chi^2_{(4)} = 1883.6, p < 0.0001$).  This linear model does not detect the
interaction of frequency by gender. By imposing linearity, the nonlinear effect
of frequency for females can only be accounted for by means of random
intercepts and slopes, but the result is a model with a substantially worse
goodness of fit.  This example illustrates an important aspect of working with
{\sc gamm}s: The model has to find the best balance between tracing variance
back to random effects and tracing variance back to wiggly curves or
(hyper)surfaces.  This is why special care is required when carrying out model
comparison, which we base on a comparison of f{\sc reml} scores using the
chi-squared test, as implemented in the {\tt compareML} function in the {\tt
itsadug} package for {\tt R}.

We have seen, both for the {\tt KKL} dataset and for the {\tt baldey} dataset,
that trial enters into significant interactions with predictors of interest.
This raises the question of whether in the absence of such interactions,
effects of trial are just a nuisance factor without theoretical interest.  We
think that even in such cases, exemplified for the {\tt baldey} dataset by the
factor smooth for session, are of more theoretical interest than one might
think.  Participants with more wiggly effects of session or trial are subjects
with more variable responses.  Ever since the study of
\citet{segalowitz1993skilled}, it is known that more skilled and automatized
processing is indexed by a lower coefficient of variation ({\sc cv}, the ratio
of the standard deviation and the mean of a subject's response times).  For the
{\tt baldey} dataset, subjects' {\sc cv} (calculated for the RTs to words, excluding
short outlier RTs) and subjects' error proportions (calculated over all trials) enter
into a strong negative correlation ($r = -0.64, t(18)=-3.6, p = 0.0021$),
indicating that subjects who are disproportionally less variable in their
reaction times are the ones who make fewer errors \citep[see
also][]{segalowitz1999rt,segalowitz2009automaticity}.  A greater variability in
reaction times could be due to greater jitter on one hand, but also to greater
fluctuations in attention and stronger effects of fatigue on the other hand.
By-subject factor smooths for trial make visible this second source of subject
variability, and are therefore diagnostic of differences in the degree to which
language processing skills have been automatized.

\subsection{The poems dataset}

Our final example addresses a data set previously discussed by
\citet{Baayen:Milin:2010}, available under the name {\tt poems} in the {\tt
RePsychLing} package.  This data set comprises a total of 275996
self-paced reading latencies from 326 subjects, for
2315 words appearing across 87
modern Dutch poems.  Words are partially nested under poems.  Any given subject
read only a subset of poems.  

Baayen and Milin included random intercepts for subject, word, and poem, as
well as several by-subject and by-word random slopes for various numerical
predictors.  These authors sought to eliminate the problem of autocorrelated
errors by including trial as a predictor, as well as the self-paced reading
latency at the preceding word.  

As discussed in detail by \citet{Bates:Kliegl:Vasishth:Baayen:2015}, the
model of Baayen and Milin is overparameterized with respect to its random
effects structure.  Given the Zipfian shape of word frequency distributions and
the large numbers of words occurring only once in the corpus of poems, data are
too sparse to include word as random-effect factor.  Furthermore, correlation
parameters for by-word random intercepts and slopes in the Baayen and Milin
model were quite large, with absolute magnitudes $> 0.8$, often an indicator of
an overparameterized model.  As for the {\tt baldey} data discussed in the
preceding section, including an {\sc ar(1)} process in the errors is a
principled and effective solution for addressing the issue of autocorrelated
errors.\footnote{
  Including the previous reaction time as covariate in order to reduce the
  autocorrelation in the error, as suggested by \citet{Baayen:Milin:2010},
  has many disadvantages compared to including an {\sc ar}(1) process in the
  errors, and is not recommended.
}
\begin{figure}
\centering
\includegraphics[width=0.8\textwidth]{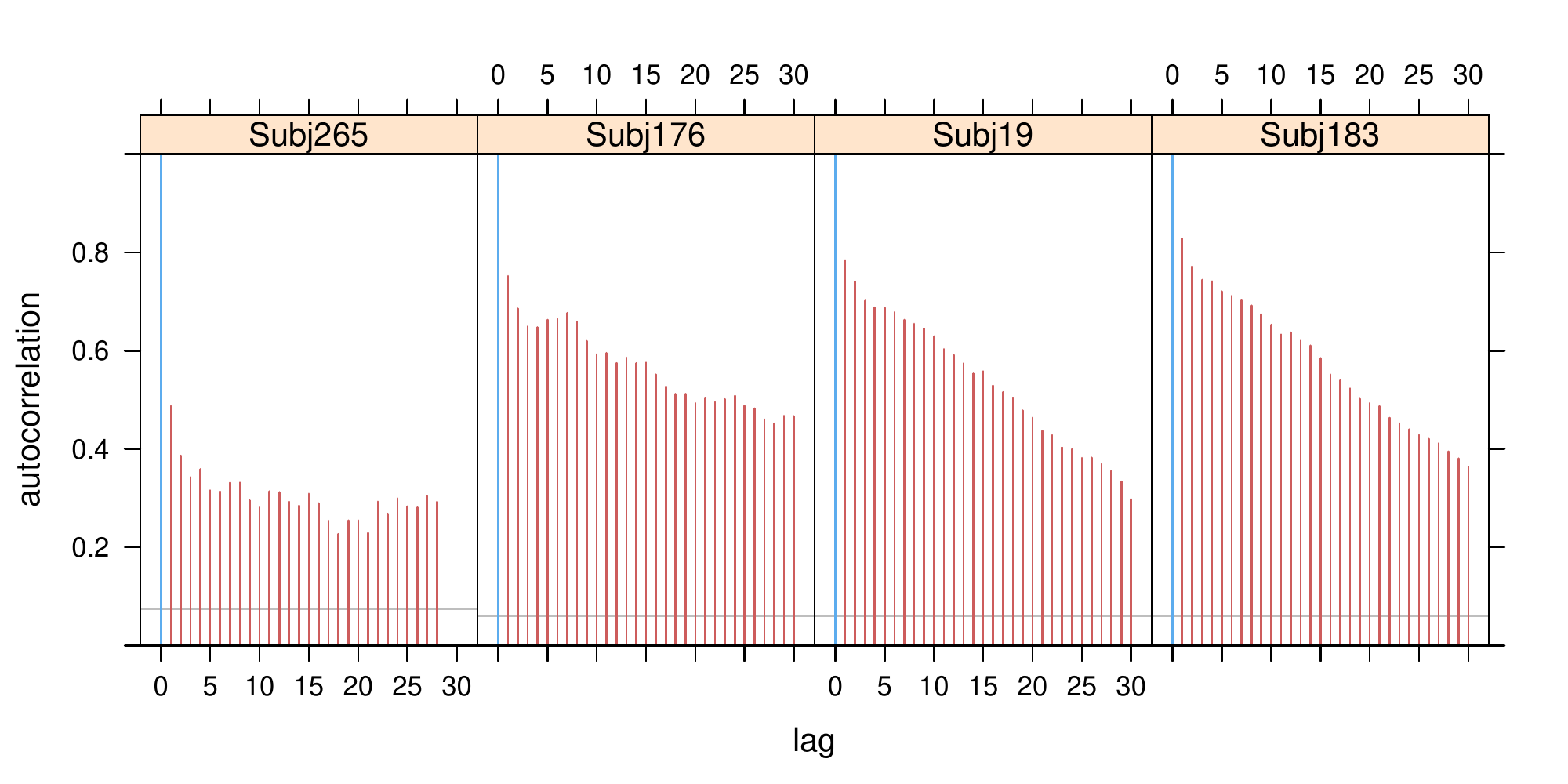} 
\includegraphics[width=0.8\textwidth]{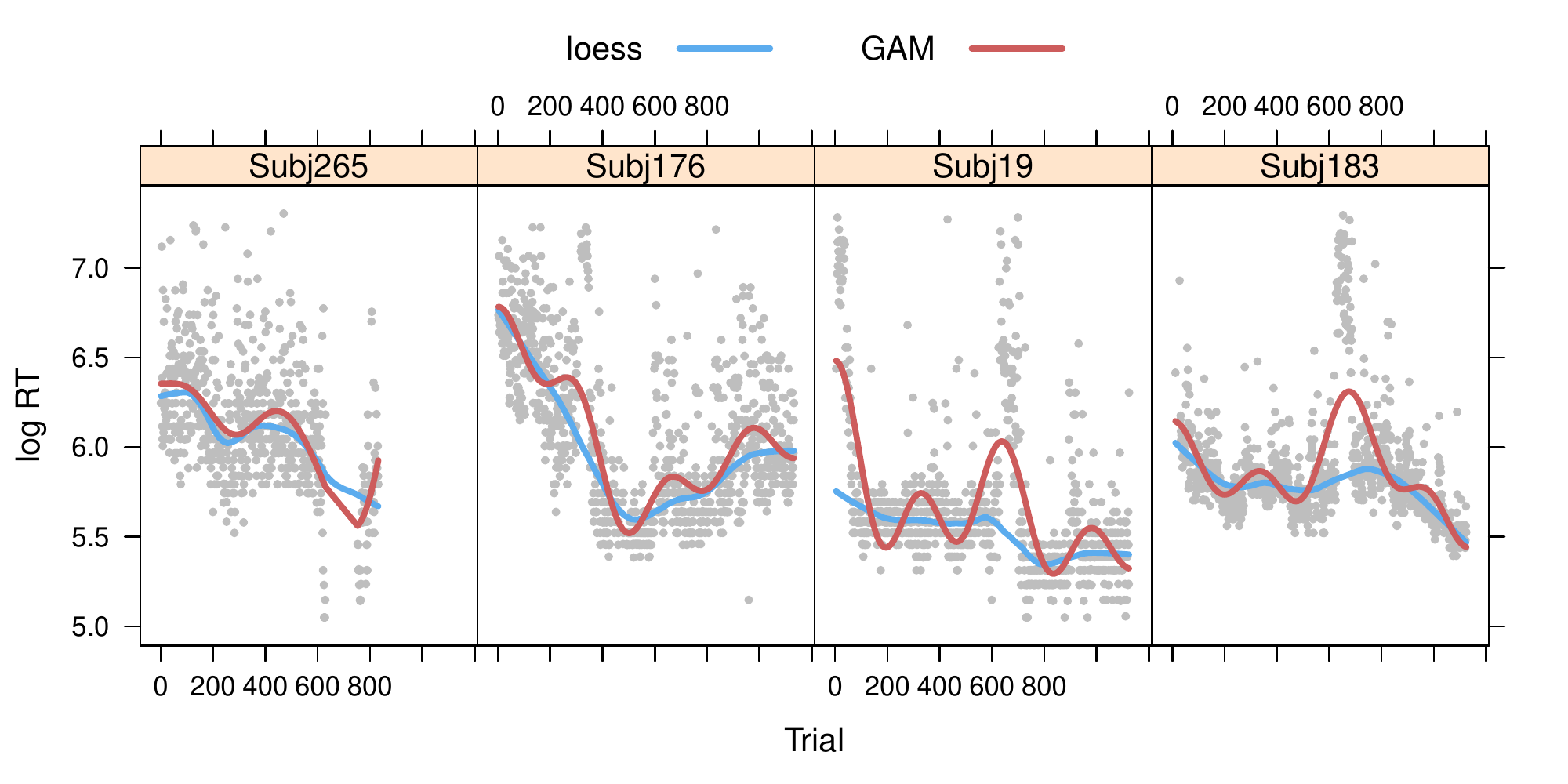}
\includegraphics[width=0.8\textwidth]{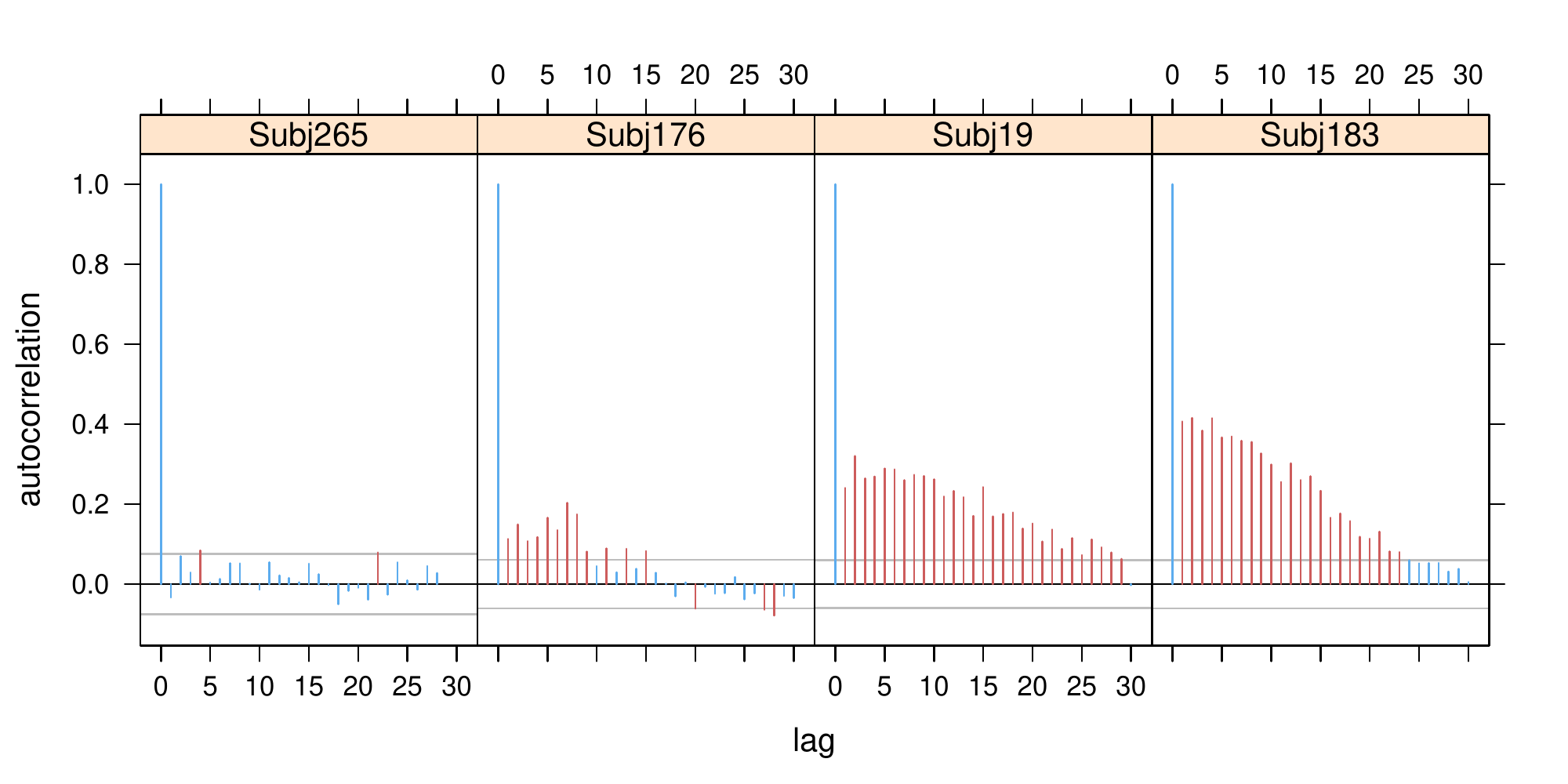} 
\caption{
Autocorrelation functions for log reaction time for four exemplary subjects in
the {\tt poems} data set (upper panel), the corresponding plots (center panels)
graphing log reaction time against trial, with a loess smoother (span = 0.2, in
blue) and a {\sc gam} factor smooth (red), and autocorrelations in the
residuals of the {\sc gamm} ($\rho = 0.3$).
}
\label{fig:poemsAuto}
\end{figure}

Within the context of the present discussion, the {\tt poems} dataset is of
interest for two reasons.  First, because subjects are reading connected
discourse rather than responding to unrelated isolated stimuli, the
autocorrelation in their responses is much stronger than in the {\tt KKL} and
{\tt baldey} datasets.  This is illustrated in the top panels of
Figure~\ref{fig:poemsAuto} for four exemplary subjects.  In this dataset, there
is only a handful of subjects without autocorrelations, and there are subjects
with even stronger autocorrelations than the ones shown here.  The second row
of panels shows the corresponding scatterplots with loess and {\sc gam}
smoothers.  Especially for subjects 19 and 183, there are temporally
concentrated spikes of long reading times that are beyond what a {\sc gam}
smooth can capture.  The lower set of panels illustrates the limitations of
what the {\sc gamm} fitted to this data set (and described in detail in
Table~\ref{tab:poetry} in the appendix) can accomplish.  For subject 265, the
autocorrelations are properly removed, and for subject 176, the reduction in
autocorrelation is perhaps satisfactory.  This is not the case for subjects 19
and 183, unsurprisingly given the spiky trends in the scatterplots.

Increasing $\rho$ is not an option.  As discussed in further detail in
\citet{Baayen:VanRij:DeCat:Wood:2015}, since different subjects typically
emerge with different degrees of autocorrelation, one would want to adjust the
$\rho$ parameter for each individual subject.  Unfortunately,  it is at present
not known how to achieve this mathematically within the framework of the
generalized additive mixed model.  As a consequence, the optimal $\rho$ is one
that strikes a balance, such that the autocorrelation for subjects with strong
autocorrelation is reduced as much as possible, without introducing artificial
negative autocorrelation at short lags for subjects with little or no actual
autocorrelation in their residuals.

\begin{figure}
\centering
\includegraphics[width=1.0\textwidth]{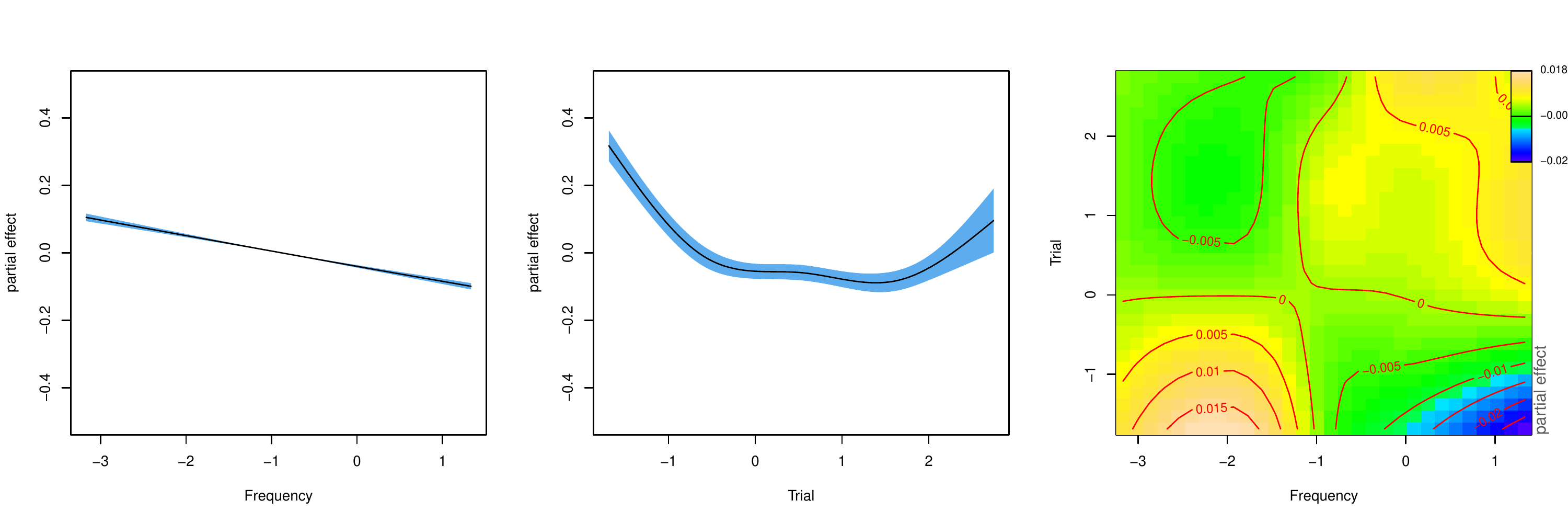}
\caption{Partial effects for {\tt Frequency} and {\tt Trial} in a {\sc gamm}
with an {\sc anova} decomposition into additive main effects and interaction,
for the {\tt poems} data set.}
\label{fig:poems.ti}
\end{figure}

Keeping in mind the caveat that the {\sc gamm} provides an imperfect window on
the complex quantitative structure of the {\tt poems} data, it is of interest
that word frequency appears to enter into a strong interaction with {\tt
Trial}. The appendix reports two models, one with a single multivariate 
smooth for these two predictors, and one in which their joint effect is
decomposed into separate, additive, main effects of {\tt Trial}, {\tt
Frequency}, and their interaction \citep[see also][]{Matuschek:Kliegl:Holschneider:2015}.  
These three partial effects are presented in Figure~\ref{fig:poems.ti}. We see
a linear facilitatory main effect of (log-transformed) {\tt Frequency} (left
panel), a U-shaped effect of {\tt Trial} (center panel), and an interaction
that rides on top of these two main effects (right panel).  The contour plot
indicates that in the early trials, frequency had a more downward-sloping
gradient. Later in the experiment, the effect of frequency is attenuated.  The
reduction in the magnitude of the frequency effect as the experiment proceeds
makes sense.  As subjects read through the poems selected for them, they tune
in to this particular genre and its vocabulary.  Words are repeated, and become
more predictable as words align into sentences, and sentences into poems.  As a
consequence, frequency of occurrence as a contextless lexical prior becomes
increasingly less informative.

\begin{figure}
\centering
\includegraphics[width=0.6\textwidth]{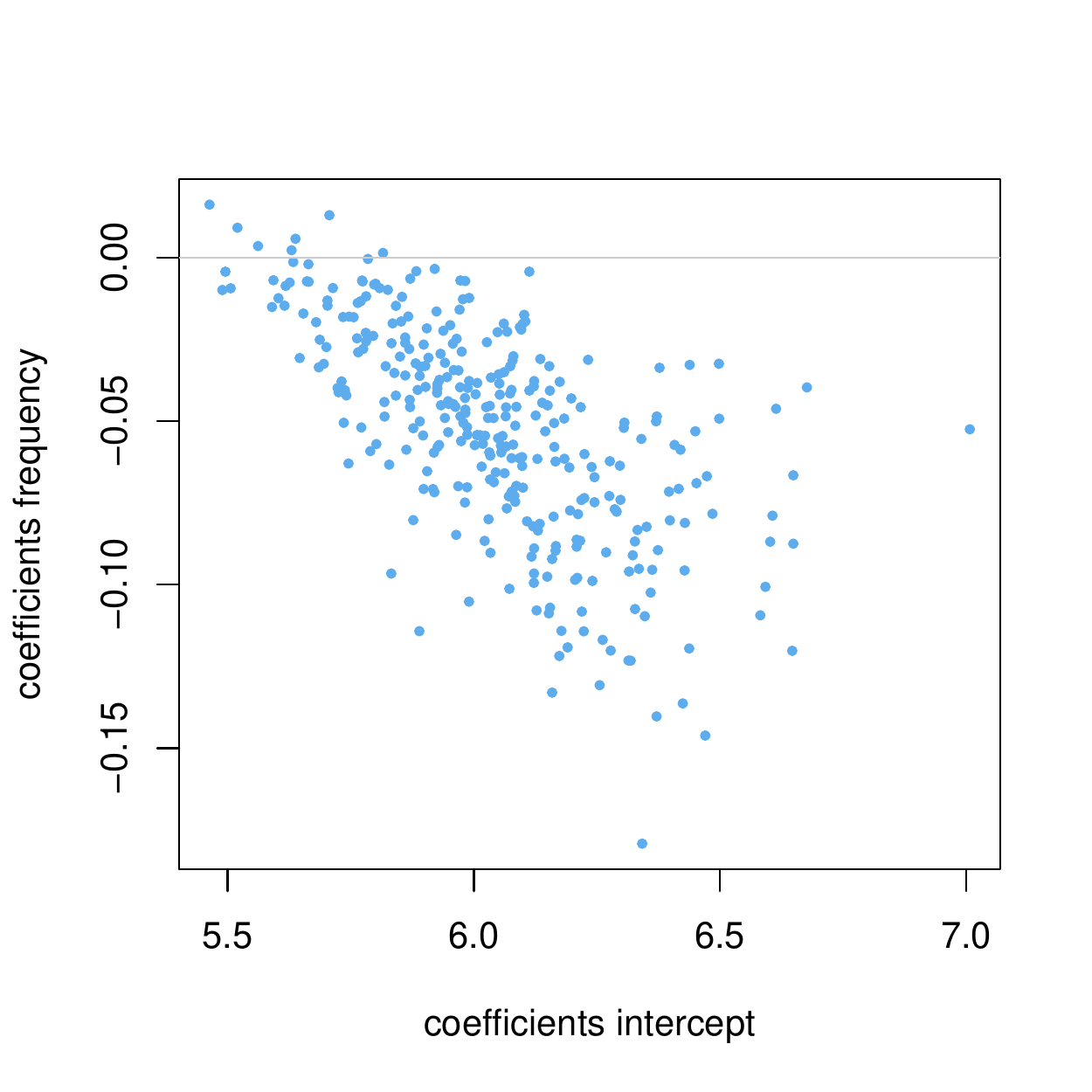}

\caption{Estimated by-subject coefficients for intercept and frequency in a
linear mixed model fitted to the self-paced reading latencies in the {\tt
poetry} data set. Slower subjects, which have larger intercepts, have more
negative coefficients for frequency, whereas the faster responders (smaller
intercepts) have coefficients for frequency close to zero.
}

\label{fig:poemsCorParam}
\end{figure}

We conclude with noting that all effects also receive generous support in a
linear mixed effects model. Although this model lacks in precision ({\sc reml}
linear model: 150493.3, {\sc reml} {\sc gamm}: 49642.31; squared correlations
of fitted and observed RTs 0.50 and 0.43 respectively), the linear mixed
model offers an insight that is not easily gleaned from the generalized
additive mixed model, namely, that the by-subject posterior modes for the
intercept and the by-subject posterior modes for frequency are negatively
correlated ($\hat{r}=-0.61$).  A correlation parameter in the linear mixed
model is well-supported by a likelihood ratio test ($p < 0.0001$).
Figure~\ref{fig:poemsCorParam} presents the by-subject coefficients (obtained
by adding the respective posterior modes to the population parameters) for
intercept and frequency. The negative correlation is well visible, and
indicates that a frequency effect is present only for those subjects who on
average respond more slowly.  This provides a further illustration that random
effects are not necessarily just `nuisance parameters', but may provide
insights that are of theoretical interest.

\section{Regression modeling strategies}

We have presented three examples demonstrating interactions of the human factor
with predictors of theoretical interest.  This raises the question of how to
proceed with the analysis of non-sterile experimental data. In what follows, we
first address this question in the context of confirmatory (or
hypothesis-testing) data analysis, and then turn to exploratory (or
hypothesis-generating) data analysis.

\subsection{Confirmatory data analysis}

An excellent introduction to confirmatory multivariate data analysis is the
monograph on regression modeling strategies by
\citet{harrell2015regression}.  For clarity of exposition, we simplify
analytical reality and describe the analysis as proceeding in three discrete
steps.  During the first step, the data are validated and explored visually,
the distributions of the response variables and the distributions of the
predictors are inspected, and transformed where necessary
\citep{Box:Cox:1964}.  In the light of what has been learned from the
initial survey of the data, including indications about non-linearities and the
potential importance of covariates and possibly the presence of the human
factor, a regression model can now be formulated.  At the second step, the
regression model is fitted to the data, and significance is assessed.  This is
the single and only time in the analytical process that a regression model is
assessed.  The third step proceeds with model criticism.  At this stage, it is
important to ascertain that the model fitted to the data at step 2 is indeed
appropriate for the data.  For a Gaussian regression model, for instance, it is
important to verify that the residuals \marked{approximately} follow a normal
distribution, that they are independently and identically distributed, and that
they do not show systematic variation with any predictors nor with the response
variable.  It is only when a critical term in a regression model withstands all
attempts to bring it down with model criticism that one may conclude that there
is reason to think that, given the simplifying assumptions that come with any
regression model (see below for further discussion), a particular effect is
actually supported.  The size of the effect, compared to the effects of other
predictors in the model, as well as the corresponding uncertainties associated
with the parameter estimates, will be essential for the assessment of the
scientific importance of this support.  Importantly, the parameters in the
model should be meaningful, at two levels.  Mathematically, parameters should
be properly estimable and interpretable.  Furthermore, at the level of domain
knowledge, all parameters should be theoretically interpretable.  For instance,
by-subject random intercepts in a regression model fitted to a reaction time
study are interpretable as a random variable placing subjects on a scale from
fast to slow responders, and by-subject factor smooths for experimental time
are interpretable as reflecting the ebb and flow of attention.

Since a confirmatory analysis allows one, and only one, statistical test for
the evaluation of a specific hypothesis, it is of crucial importance that this
test is based on models that are not too complex to be estimable, on models
that are properly interpretable, and on models that take the human factor into
account if it is present.   How then might one proceed under these stringent
boundary conditions?

At first sight, it might be argued that a model should be fit to the data that
is as complex as possible, a model that takes all possible contingencies into
account that might put the critical model parameter in jeopardy.  Thus, one
might think that it is straightforward to enrich a maximal linear mixed model
with predictors targeting the human factor.  Unfortunately, once one enters the
nonlinear world, this is even less advisable than for the linear world,  for a
variety of reasons.

First, more elaborate models can quickly become very difficult to understand.
By way of example, a model with a four-way interaction of {\tt Word Duration},
{\tt Session}, {\tt Frequency} and {\tt Trial} improves substantially on the
model presented above for the {\tt baldey} data set ($\chi^2_{9}=24.35, p <
0.0001$).  But what we learn from this four-way tensor product is unclear.  In
1959, Sigmund Koch wrote that ``Psychology [is] unique in the extent to which
its institutionalization preceded its content and its methods preceded its
problems.'' \citep[][p.~783]{koch1959psychology}. Whereas this may not be
true for all areas of psychological science (e.g., some areas of vision
research), it certainly applies to the domain of lexical processing. Here, {\sc
gamm}s will often be informative about possible structure in experimental data
that is far beyond what current theories can explain or predict.  For the {\tt
baldey} data, we deliberately avoid a `maximal' model, as, given current
knowledge, it is unclear whether such a model would contribute to understanding
the data. 

Second, when we make use of a factor smooth with shrinkage to fit nonlinear
by-subject trends over experimental time, we are making many simplifying
assumptions, among which (i) that all subject smooths can be captured with
the same smoothing parameter, and (ii) that these temporal trends do not
interact with other predictors in the model, whether factorial (say, a priming
condition) or metric (say, frequency or valence).  These assumptions may or may
not be valid, but it typically does not make much sense to aim for a complex
model term such as a tensor product smooth for trial by frequency by valence by
priming by subject, with or without shrinkage.  Again, given our current state
of knowledge, such complex models, if at all estimable, will typically be very
difficult to interpret.

Third, fitting a complex generalized additive mixed model is not a trivial
issue, and for results to be sensible, it is crucial to avoid random effects
structure that is internally collinear \citep[see][for detailed
discussion]{Bates:Kliegl:Vasishth:Baayen:2015}.  In general, as observed by
Wood (documentation for {\tt gam.selection} in the {\tt mgcv} package),
\begin{quote}
The more thought is given to appropriate model structure up front, the more
successful model selection is likely to be. Simply starting with a hugely
flexible model with `everything in' and hoping that automatic selection will
find the right structure is not often successful. 
\end{quote}

Would researchers come to `wrong' conclusions if they analyze data simply using
maximal linear models, without taking the human factor into account, without
paying attention to whether the model is overfitting the data with
mathematically uninterpretable parameters
\citep{lele2012estimability,Bates:Kliegl:Vasishth:Baayen:2015}, and
accepting less than nominal power \citep{Matuschek:2015}?  
The problem here is that low-hanging fruit is easily plucked,
often by simple linear models without any random effects.  The devil is in the
details.  Significance of factorial contrasts may not change, or may not
change by much, when the human factor is taken into account.  For the {\tt
KKL} data set, we showed that a full mixed model would lead to the conclusion
that the main effect of {\tt Orientation} is not significant, whereas a model
that takes the human factor into account suggests otherwise.  Furthermore, the
maximal model suggests that the interaction of {\tt Size} and {\tt Orientation}
is significant, but the {\sc gamm} with predictors for the human factor
indicates that there is no support whatsoever for such an interaction.
Details change, but the three-way interaction of {\tt Orientation} by {\tt Size}
by {\tt Gravitation} remains.  Do the details matter?

If details don't matter, in many cases the analyst will be well off with a
simple linear model, even a linear model without random effects.  {\color{black}
Often, conclusions about `significance' do not change when data sets are
analysed with much simpler models.}  However, when a
simple linear model produces the same verdict on significance as a linear mixed
model, or a linear mixed model provides the same verdict of significance as a
generalized additive mixed model, this does not mean that the more complex
modeling technique is not required.  Even when conclusions about significance
of predictors do not change for observed examples, they might change
substantially for as yet unobserved examples.  More importantly, random slopes
and random intercepts typically modulate effect sizes and degrees of
uncertainty.   Especially when it comes to prediction, more precise estimates
that take into account subject and item variability are invaluable.  Similarly,
taking into account nonlinearities and human factors may modulate conclusions
about effect sizes and the precise nature of functional relations.  For the
{\tt KKL} dataset, we observed a significant partial effect of {\tt
Orientation}, modulated by interactions with far smaller effect sizes.  The
partial effect of {\tt Orientation} is of theoretical significance, and it
therefore is important to use a modeling strategy that makes its partial effect
visible.\footnote{
  It might be argued we have not shown that addressing autocorrelated errors
  changes conclusions about the effect of {\tt Orientation}.  The argument runs
  as follows. The main effect of a predictor $X$ that interacts with a
  predictor $Y$ specifies the effect of $X$ when $Y=0$.  Since {\tt
  Orientation} interacts with {\tt Trial}, the main effect of {\tt Orientation}
  specifies its effect when {\tt Trial} = 0 (i.e., in the middle of the
  experiment, since {\tt Trial} was scaled). From this, it would follow that we
  have no case to argue that it is the explicit treatment of autocorrelated
  errors that has changed the apparent conclusions from the model about the
  effect of {\tt Orientation}.  This argument misses three important points.

  First, autocorrelation in the errors is addressed in part by the by-subject
  factor smooths for {\tt Trial}, and in part by the autocorrelation parameter
  $\rho$.  It is the combination of the two that leads to different conclusions
  about the effect of {\tt Orientation}.  Second, as explained in
  section~\ref{sec:gamms}, main effects in a model with interactions are
  crucial for properly calibrating the wiggly curves for individual factor
  levels with respect to the intercept.  They are an essential part of the
  model.  Changing orientation from cardinal to diagonal implies a modulation
  of the intercept by 0.078 units on the -1000/RT scale.  This change effects
  all trials for the relevant factor level.   A maximal {\sc lmm} estimates the
  effect to be much smaller (0.028) and not significant.  In other words, the
  maximal {\sc lmm} underestimates the distance between the relevant curves.
  Third, as a consequence, the maximal {\sc lmm} provides a warped perspective
  of the magnitude of the effect of {\tt Orientation} vis-\`{a}-vis the other
  predictors in the model.  
}

Given that a maximal model approach provides the false security of a comfort
blanket, the question remains how one might proceed under the stringent
boundary conditions of confirmatory data analysis.  All we can do to answer
this question is present examples of how one might proceed.  Consider a
chronometric experiment with subjects and items.  Inspired by
\citet{harrell2015regression}, one possible way to proceed could be as follows.
As a first step,
following data validation, exploratory visualization is carried out.  At this
stage, non-parametric scatterplot smoothers could be used to probe for the
presence of by-subject trends over experimental time.  Furthermore, the
autocorrelation function could be obtained for the response variable, in order
to assess what value of $\rho$ might be required.  However, because the
temporal autocorrelation can be due to the combined presence of an {\sc ar}(1)
process in the errors and subject-specific trends in experimental time, the
autocorrelation function for the response may overestimate the value of $\rho$
when by-subject trends in experimental time are in fact present.  Therefore, it
may be preferable to fit an intercept-only {\sc gamm} with factor smooths for
subject and random intercepts for item to the data, with as only aim to detect
with more precision the extent to which the human factor is present, to
determine whether it is necessary to include by-subject factor smooths, and to
obtain an estimate for the autocorrelation parameter $\rho$.  If there is no
clear evidence for the human factor, a linear mixed model is an excellent
choice, otherwise, a {\sc gamm} is preferable, with an autocorrelation
parameter for the {\sc ar(1)} process in the errors set close to the
autocorrelation at lag 1 observed for the intercept-only model.  

Then, at step two of the analysis,  a model could be fitted with all relevant
fixed-effect parameters added in, but without any further random effects and
random slopes.  Significance of the critical predictor can now be assessed
through model comparison with a simpler model from which the critical
predictor, or the relevant critical interaction with this predictor, is
removed.  Importantly, this is the single and only time at which significance
is assessed.  

The final step proceeds to model criticism.  At this step, the question is
whether significance (if established) will survive removal of overly
influential outliers, addition of further random-effect parameters
(specifically, and importantly, random effects or random slopes for the
critical model terms), adjustment of the autocorrelation parameter if
necessary, and inclusion of interactions with human-factor variables.
Bootstrap validation is also worth considering at this step.  If an effect
withstands model criticism, it can be reported as significant with the p-value
obtained at step 2, otherwise, it should be reported as not significant.   This
confirmatory modeling strategy has the advantage that models that overfit the
data with meaningless parameters are avoided.  As pointed out by
\citet{lele2012estimability}, ``Whenever mixed models are used, estimability
of the parameters should be checked before drawing scientific inferences or
making management decisions''.   Since meaningless parameters can arise even
under convergence \citep[see][]{Bates:Kliegl:Vasishth:Baayen:2015}, the
analyst may want to minimize the risk of running into this situation
specifically when significance is assessed in a confirmatory context. 

Importantly, there are other strategies that could be followed, such as
starting with a model including all random intercepts and all random effects
and slopes, while leaving out correlation parameters
\citep{Bates:Kliegl:Vasishth:Baayen:2015}.  Here, a confirmatory
setting takes the significance of the pertinent predictor as the outcome of
interest, and subsequent model criticism is carried out to ensure that this
significance is trustworthy.  If it turns out that the model is too complex to
be supported by the data, the analyst may want to refit a simpler and better
justified model, in which case the analysis has become exploratory --- it is
only as long as significance is evaluated once, and once only, with subsequent
model criticism to ensure support for the original significance test, that the
analysis is a proper confirmatory analysis.  What specific strategy is
followed, and we have given only two of many possible strategies, is, to a
large extent, a matter of taste, so it would make sense to report the details
of the strategy that was actually followed for a given confirmatory analysis.
\marked{In any case, the most transparent way to proceed is to release all 
data and code with the published paper, so that readers have the option to
draw their own conclusions from the data.}

\subsection{Exploratory data analysis}

One of the causes of the deplorable rate of replication failure for psychology
(and many other fields of inquiry) is that confirmatory data analysis is seen
as vastly superior to exploratory data analysis, and that as a consequence, the
results of many exploratory data analyses are presented as if they were the
result of confirmatory analysis.  

All models reported in the present study are the result of data exploration
with step-by-step theory-driven incremental model building \citep[for
examples of such an approach, see][]{Pinheiro:Bates:2000,gonzalez2014linear}.
The t and p-values reported in the appendix, therefore, are indicators of
surprise and should not be taken at face value as exact probabilities.  (Note,
however, that the crucial probabilities for the human factor are so small,
often $<$2$e$-16, that they may be expected to survive substantial correction
for multiplicity.) We believe that exploratory data analysis is of great
importance for those domains of inquiry where explicit and mathematically
precise theories are lacking.  In these areas of inquiry, results of
experiments can be completely opposite to what was expected, even though in
hindsight, they may make perfect sense \citep[see, e.g., the anti-frequency
effect in Figure~\ref{fig:vietnamese}, for a computational model based on
discriminative learning that captures this effect,
see][]{Pham:Baayen:2015}.\footnote{The review process forced a presentational
style on this study in which the experimental results are reported as being
predicted by discriminative learning theory, whereas the unexpected
experimental results preceded in time the subsequent modeling with
discriminative learning.} Importantly, in an exploratory setting, one can
actually learn from the data, in a multivariate setting often in many
dimensions simultaneously, instead of receiving only confirmation (or
disconfirmation) of a single hypothesis.

Also in an exploratory setting, model criticism is absolutely essential.  An
effect is worth taking seriously only if it withstands truly serious attempts
to bring it down.  Researchers may want to complement exploratory regression
analysis with techniques from machine learning such as random forests
\citep{Breiman:2001,Strobl:Malley:Tutz:2009} or gradient boosting machines
\citep{Friedman:2001,ChenXBoost:2015} to obtain an independent assessment
of variable importance that is orthogonal to the exploration of the data with
regression modeling.  These techniques are not plagued by issues of
collinearity \citep{Wurm:Fisicaro:2014}, and they do not require any kind
of model selection \citep[for an example, see, e.g.,][for application of
random forests for this purpose]{Baayen:Milin:Ramscar:2016}.

\subsection{A gold standard?}

The \citet{open2015estimating} reported the results of an extensive series of
replication studies, and documented a 50\% drop in observed effect sizes and a
drop from 97\% to 36\% of significant results.   Publication bias
\citep{francis2012publication}, non-random selection of stimuli and subjects
\citep{Forster:2000,sander2006rogue,Francis:2013}, societal changes
\citep{Ramscar:Shaoul:Baayen:2015} and especially lack of power
\citep{button2013power,Westfall:Kenny:Judd:2014} are major concerns.  Clearly,
many of the results in the published literature are but moving shadows of the
true effects, which raises the question how to escape from the current
methodological cave.

\citet{barr2013random} proposed a new standard for the statistical analysis
of experimental data.  Based on a series of simulation studies, they argued
that anti-conservative p-values are best avoided by fitting linear mixed
effects models with all variance components included that could in principle be
non-zero given the experimental design.  The simulations on which
\citet{barr2013random} base their recommendation make very specific
assumptions:
\begin{quote}
  We assumed no list effect; that is, the particular configuration of `items'
  within a list did not have any unique effect over and above the item effects
  for that list. We also assumed no order effects, nor any effects of practice
  or fatigue. (p. 264)
\end{quote}
Given our limited understanding of the human factor, this simplification is
understandable.  However, this very simplification invalidates their simulation
design as a foundation for a general gold standard.  We have shown that the
human factor may interact with key predictors, and that it may lead to
different conclusions about the details of their effects.  


Since the proposed standard of Barr et al. is overly conservative with an
unnecessary loss of power \citep{Matuschek:2015}, and since it comes with the
risk of basing conclusions on mathematically ill-defined models that overfit
the data \citep{lele2012estimability,Bates:Kliegl:Vasishth:Baayen:2015}, Barr
et al.'s proposal has the unfortunate side-effect of locking analytical
practice in a methodological cage of shadows from which the true structure of
experimental data, rich and fertile in perhaps unexpected ways, cannot be fully
appreciated.

{\color{black} Recommendations such as the one of Barr et al., which once in
the literature quickly rise to the status of rules enforced in the review
process with an iron fist, have the unfortunate side effect of diverting
attention from the model, the balance of forces within the model, the
uncertainties associated with the model, and its inevitable weaknesses, to the
so fervently desired but so over-valued p-value.  A one-size-fits-all rule for
obtaining such p-values might seem attractive as the only way to obtain an
`objective' procedure evaluating experimental effects. However,} irrespective
of whether analysis is exploratory or confirmatory, in the modern age,
objectivity can be achieved in a much more direct way.  Whereas before the
advent of the internet, reporting p-values in printed journals was the only way
to make an argument that a particular effect is present, current communication
technology makes it possible to publish not only p-values but the data
themselves, using platforms such as the Mind Research Repository at
\url{http://openscience.uni-leipzig.de/index.php/mr2} or the Open Science
Framework (\url{https://osf.io/}), where data can be made available together
with details on the statistical analysis.  With the data out in the open,
readers will not only be able to evaluate for themselves the appropriateness of
the analyses reported in a journal, but opportunities are created for improved
analyses, either with current or with future statistical software.  In this
way, a degree of objectivity can be reached that goes far beyond what can be
obtained by mechanical procedures and the considerable risk of associated
artifacts.  Furthermore, through meta-analysis, one can build on previous
findings.

\section{Discussion}

Even though there is considerable awareness in the literature on memory and
language that time series of behavioral responses are not independent
\citep{Broadbent:1971,Welford:1980,Sanders:1998,Taylor:Lupker:2001,deVaan:Schreuder:Baayen:2007,Baayen:Milin:2010,masson2013modulation},
the fact that this interdependence has far-reaching consequences for the
statistical analysis of experimental data has not received systematic
reflection.  We have reported three data sets in which inter-trial dependencies
are present.  Unlike molecules or barley, the units from which response
variables in psychology are harvested are intelligent beings that constantly
adapt to their environment.  Humans learn.  They get tired.  Their attention
wanes, and then is refocused on the task.  When attentional and adaptative
processes are demonstrably present in experimental data, and interact with
predictors of theoretical interest, it is advisable to bring these processes
under statistical control.  Failure to take the human factor into consideration
comes with the risk of misunderstanding the finer details of the quantitative
structure of the data and the extent to which this structure is shaped by the
predictors of interest.

We have introduced the generalized additive mixed model as an extension of the
linear mixed model that makes it possible to bring the human factor into the
statistical model, and to take the statistical sting out of the
autocorrelations in the residual error.  We do not wish to claim that with {\sc
gamm}s researchers will finally emerge from the cave of shadows and apprehend
the true effects themselves.  But we are finding {\sc gamm}s  helpful in
sharpening the contours and outlines of these effects.  For almost all data
sets that we have investigated with {\sc gamm}s, we have obtained better fits
by including by-subject factor smooths for trial.  We also demonstrated
interactions of trial with predictors of interest, a first step towards a
better understanding of the human factor in lexical processing.

As any statistical model, {\sc gamm}s build on assumptions that are hoped to be
reasonable, but of which we often know they involve substantial
simplifications. When using {\sc gamm}s, it is important to have good coverage
of the covariate space, especially when using generalized models with Poisson
or binomial families, and highly irregular regression surfaces with highly
localized effects should be treated with caution.\footnote{
{\sc Gam}s do not require especially large data sets. Examples from
\citet{Wood:2006} include a data set with only 31 observations on girth,
height, and volume for black cherry trees, where a thin plate regression spline
for girth appears justified, and a data set with 634 observations on mackerel
eggs where a thin plate regression spline for longitude and latitude is part of
the model specification.  Of course, when there is only a handful of distinct
values for a given covariate, a smooth for that covariate will not make sense.
The smoothers in the {\tt mgcv} package typically produce an error message for
such cases.} 
Unlike the linear mixed model, the {\sc gamm} as implemented in the {\tt mgcv}
package does not offer the possibility to test for correlation parameters in
the random effects.  Analysts used to the speed with which the software of the
fourth author fits {\sc lmm}s will find the speed with which {\sc gamm}s with
complex random effects structure converge excruciatingly slow, the price paid
for not imposing prior constraints on the random effects structure.
Furthermore, the penalized factor smooths that we have used assume a common
smoothing parameter for all subjects, which may be true,  but may also be
incorrect.  The factor smooths build on spline theory, but when time series of
reaction times are spiky instead of smoothly wiggly, splines are better than
nothing, but certainly not perfect.  To this list of limitations we can also
add that the adjusting of the errors for an {\sc ar}(1) autocorrelative process
will often be too simplistic in two ways.  First, as we have demonstrated,
$\rho$ should ideally vary with participant, which is currently not possible
with the {\tt gam} (and {\tt bam}) functions of the {\sc mgcv} package (but
progress may be possible here by going fully Bayesian, see Wood, 2016).
Second, there is no guarantee that the autocorrelative process is a simple {\sc
ar(1)} process --- dependencies may well stretch further back in time.
Nevertheless, we believe that with {\sc gamm}s, researchers have a tool in hand
with which the real-life complexities with which psycholinguistic data may be
infected can start to be investigated.

We conclude with some reflections on parsimony in regression modeling.
George Box is well known for stating that all models are wrong, but that
some are useful \citep{Box:1979}.  With respect to model parsimony,
he noticed that
\begin{quote}
Since all models are wrong the scientist cannot obtain a ``correct'' one by
excessive elaboration. On the contrary following William of Occam he should
seek an economical description of natural phenomena. Just as the ability to
devise simple but evocative models is the signature of the great scientist so
overelaboration and overparameterization is often the mark of mediocrity. 
\citep[][p.\ 792]{Box:1976}
\end{quote}
When working with {\sc gamm}s, models of considerable complexity can be fit to
experimental data sets.  Since language is a complex cognitive skill serving
users interacting in highly complex and technologically advanced societies, it
is perhaps not surprising that statistical models may need to be more complex
than previously thought when it comes to taking into account the human factor.
Within the boundary conditions of not overparameterizing, of properly balancing
Type~I and Type~II error rates, and of having a model with meaningful and
interpretable parameters, the challenge is to find the right balance between
model simplicity and faithfulness to the data. 

Within the context of confirmatory data analysis, a possible modeling strategy
might be to keep the model lean, with factor smooths and an autocorrelation
parameter included when there appears to be evidence for the human factor, and
to shift the burden of securing a reasonable degree of confidence about a
critical effect to model criticism.  It must be acknowledged that an important
achievement of the \citet{barr2013random} paper was to raise awareness about
the dangers of fitting overly simple linear mixed models. However, their
recommendation (summarized in the title  of their paper) to ``keep it
maximal'', crosses over to such an extreme position as to be untenable in most
realistic statistical modeling settings.

Within the context of exploratory data analysis,  incremental hypothesis-driven
model building can yield an accumulation of valuable insights.  Importantly,
theory and experience should guide model building, counterbalancing
faithfulness to the data with a drive for simplicity.  For the {\tt KKL} data
set, for instance, the three-way interaction of {\tt Orientation} by {\tt Size}
by {\tt Trial} was explored because we thought it was conceivable that
within-experiment learning and adaptation might vary with the difficulty of the
different testing conditions.  It is possible that the details of this
interaction vary from subject to subject.  Without hypotheses about what might
drive individual differences in a four-way interaction with subject, such an
interaction, even if it were estimable, would be extremely difficult to
understand.  For the {\tt baldey} data set, we considered the possibility that
listeners might adapt to the speech rate of the speaker as the experiment
proceeded.  However, we refrained from discussing a model in which {\tt Trial}
and {\tt Word Duration} entered into further interactions with {\tt Session}
and {\tt Frequency}, even though such a four-way interaction, which is
estimable, improves substantially on the model presented above for the {\tt
baldey} data.  This more complex model might be capturing something real, but
without guidance from theory, and without further support from replication
studies, it is unclear what the benefits of such a complex model might be.  For
the {\tt poetry} data set, the model reported in \citet{Baayen:Milin:2010}
turned out to be overparameterized, the data for words being too sparse to
allow inclusion of by-word random intercepts and slopes. Therefore, our {\sc
gamm} for the {\tt poetry} data did not include random effects for words.   In
short, also in the context of exploratory data analysis, parsimony is a virtue,
not a vice.

The probability of objectivity and replicability in data analysis, whether
exploratory or confirmatory, is likely to be enhanced when researchers make
their data and analyses available to the research community.  Of course, making
data available does not protect against experimental fishing expeditions, such
as running additional subjects until significance is obtained, or running a
pilot study and selecting for the main experiment that subset of items that in
the pilot study yielded the desired effect.  Transparency in reportage is
therefore also of great importance. However, just by itself, making the
data available is a substantial deterrent for reporting results of fishing
expeditions: Anyone with access to the data will immediately spot that the
model published is an implausible one.  Publication of data will also render
fishing expeditions across multiple methods for analyzing the data less
attractive.  For instance, reporting an F1+F2 analysis
\citep{Forster:Dickinson:1976} that supports significance after first having
observed a lack of significance in a {\sc lmm} is straightforward to detect.
Rather than seeking to guarantee objectivity through rigid methodological
one-size-fits-all prescriptions, we think open science, combined with
responsible statistical analysis, is the way forward out of the cave of
shadows.


\theendnotes


\bibliography{/home/harald/atransfer/data}

\ \\
\ \\
\noindent
{\bf Acknowledgements} \\
\ \\
This research was supported by an Alexander von Humboldt Professorship awarded
to the first author.  {\color{black} We are indebted to Simon Wood for
invaluable discussion and feedback and for the many extensions he added to the
{\tt mgcv} package to facilitate the analysis of (psycho)linguistic data.}

\newpage

\appendix
\begin{flushleft}
  {\Large {\bf Appendix}}
\end{flushleft}

\begin{center}
{\bf The KKL dataset}
\end{center}

\definecolor{Gray}{gray}{0.9}

\begin{table}[ht]
\centering
{\footnotesize

{\bf random effects} \\ \ \\
\begin{tabular}{llrr} \hline
 Groups               & Name       & Std.Dev.   &   Corr    \\ \hline
 subj                 & Intercept  & 0.161766   &           \\ \hline
 \rowcolor{Gray}      & Trial      & 0.051180   & -0.273    \\ \hline
 subj.1               & spt        & 0.066809   &           \\
 subj.2               & grv        & 0.033814   &           \\
 subj.3               & obj        & 0.025609   &           \\
 subj.4               & orn        & 0.076247   &           \\
 subj.5               & spt\_orn   & 0.033226   &           \\ \hline
 Residual             &            & 0.185591   &           \\ \hline
\end{tabular}

\ \\
\ \\
{\bf fixed effects} \\ \ \\
\begin{tabular}{lrrr} \hline
              & Estimate & Std. Error & $t$ value \\ \hline
Intercept     &  5.659 & 0.018 & 322.91 \\ 
  sze         &  0.184 & 0.035 &   5.27 \\ 
  spt         &  0.074 & 0.008 &   9.62 \\ 
  obj         &  0.043 & 0.005 &   9.41 \\ 
  grv         & -0.001 & 0.005 &  -0.17 \\ 
  orn         &  0.014 & 0.009 &   1.51 \\ 
  Soa L       & -0.010 & 0.001 & -12.56 \\ 
  Soa Q       &  0.019 & 0.001 &  20.61 \\ 
  sze:spt     &  0.048 & 0.015 &   3.14 \\ 
  sze:obj     & -0.012 & 0.009 &  -1.30 \\ 
  sze:grv     & -0.037 & 0.010 &  -3.66 \\ 
  sze:orn     &  0.039 & 0.018 &   2.20 \\ 
  spt:orn     &  0.020 & 0.006 &   3.12 \\ 
  obj:orn     &  0.009 & 0.007 &   1.26 \\ 
  grv:orn     &  0.011 & 0.007 &   1.52 \\ 
  sze:spt:orn & -0.014 & 0.013 &  -1.10 \\ 
  sze:obj:orn & -0.003 & 0.014 &  -0.24 \\ 
  sze:grv:orn & -0.047 & 0.014 &  -3.25 \\ \hline
  \rowcolor{Gray} Trial L     & -0.043 & 0.006 &  -7.47 \\ 
  \rowcolor{Gray} Trial Q     &  0.015 & 0.001 &  16.96 \\ 
  \rowcolor{Gray} sze:Trial   &  0.018 & 0.011 &   1.60 \\ 
  \rowcolor{Gray} orn:Trial   &  0.028 & 0.003 &   8.64 \\ 
  \rowcolor{Gray} sze:TrialQ  & -0.000 & 0.002 &  -0.05 \\ 
  \rowcolor{Gray} orn:TrialQ  & -0.006 & 0.005 &  -1.27 \\  \hline
\end{tabular}
}
\caption{Summary of the LMM fitted to the KKL data set. Extensions to the reference model are highlighted.
For factors, $-$0.5/$+$05 dummy coding was used. L: linear, Q: quadratic.}

\label{tab:KKLlmer3}
\end{table}

\clearpage

\begin{table}
\centering
{\footnotesize
\begin{tabular}{lrrrr} \hline
A. parametric coefficients & Estimate & Std. Error & t-value & p-value \\  \hline
  Intercept & 5.6851 & 0.0189 & 300.4450 & $<$ 0.0001 \\ 
  sze & 0.1840 & 0.0378 & 4.8608 & $<$ 0.0001 \\ 
  spt & 0.0729 & 0.0079 & 9.2050 & $<$ 0.0001 \\ 
  obj & 0.0411 & 0.0041 & 9.9844 & $<$ 0.0001 \\ 
  grv & -0.0005 & 0.0049 & -0.1003 & 0.9201 \\ 
  orn & 0.0375 & 0.0142 & 2.6441 & 0.0082 \\ 
  sze:spt & 0.0483 & 0.0158 & 3.0453 & 0.0023 \\ 
  sze:obj & -0.0088 & 0.0082 & -1.0749 & 0.2824 \\ 
  sze:grv & -0.0366 & 0.0099 & -3.7074 & 0.0002 \\ 
  sze:orn & 0.0096 & 0.0283 & 0.3375 & 0.7357 \\ 
  spt:orn & 0.0213 & 0.0064 & 3.3449 & 0.0008 \\ 
  obj:orn & 0.0083 & 0.0068 & 1.2129 & 0.2252 \\ 
  grv:orn & 0.0078 & 0.0068 & 1.1387 & 0.2548 \\ 
  sze:spt:orn & -0.0098 & 0.0128 & -0.7668 & 0.4432 \\ 
  sze:obj:orn & -0.0075 & 0.0136 & -0.5502 & 0.5822 \\ 
  sze:grv:orn & -0.0483 & 0.0137 & -3.5241 & 0.0004 \\ \\ \hline \hline
B. smooth terms & edf & Ref.df & F-value & p-value \\ \hline
\rowcolor{Gray} fs(Trial,subj)             & 557.8665 & 774.0000 & 661.8472 & $<$ 0.0001 \\
  re(subj,spt)                             &  77.7931 & 86.0000 & 31.5911 & $<$ 0.0001 \\ 
  re(subj,grv)                             &  47.6056 & 84.0000 & 2.1643 & $<$ 0.0001 \\  
  re(subj,obj)                             &  29.5712 & 84.0000 & 1.3404 & 0.0011 \\      
  re(subj,orn)                             &  44.0796 & 84.0000 & 1.1481 & $<$ 0.0001 \\  
  re(subj,spt\_orn)                        &  47.3734 & 84.0000 & 1.3153 & $<$ 0.0001 \\  
  \rowcolor{Gray} s(Trial): big+cardinal   &  8.3489 & 8.6794 & 13.4016 & $<$ 0.0001 \\   
  \rowcolor{Gray} s(Trial): small+cardinal &  8.0534 & 8.4887 & 9.8373 & $<$ 0.0001 \\    
  \rowcolor{Gray} s(Trial): big+diagonal   &  5.8875 & 6.5738 & 5.8564 & $<$ 0.0001 \\    
  \rowcolor{Gray} s(Trial): small+diagonal &  7.9974 & 8.4525 & 8.8624 & $<$ 0.0001 \\    
  s(Soa)                                   &  5.6017 & 6.7468 & 103.1147 & $<$ 0.0001 \\  \hline
\end{tabular}
}
\caption{
Summary of the full {\sc gamm} fitted to the {\tt KKL} data set.  Extensions to
the reference model are highlighted. For factors, $-$0.5/$+$05 dummy coding was
used.  A separate factor was defined with four levels, one for each combination
of {\tt Size} and {\tt Orientation}, and a thin plate regression spline ({\tt
s()}) was fitted for each of its four levels.  {\tt re(X,Y)} denotes random
intercepts in $Y$ for grouping factor $X$.  The penalized factor smooth for
subject ({\tt fs(Trial, subj)} includes by-subject intercept calibration.
}
\label{tab:KKLfullgam}
\end{table}

\clearpage

\begin{center}
{\bf The baldey dataset}
\end{center}

\begin{table}[ht]
\centering
{\footnotesize
\begin{tabular}{lrrrr} \hline
                                       & Estimate & Std. Error &  t value & Pr($>$$|$t$|$) \\ \hline
Intercept                              &    -0.96 &       0.04 &   -22.47 &  $<$ 0.0001    \\
gender=male                            &     0.21 &       0.06 &     3.49 &  $<$ 0.0001    \\ \hline\hline
                                       &      edf &     Ref.df &        F &     p-value    \\ \hline
s(LemmaFreq):gender=female             &     3.20 &       3.61 &     53.51&  $<$ 0.0001    \\ 
s(LemmaFreq):gender=male               &     2.26 &       2.51 &     52.23&  $<$ 0.0001    \\ 
te(word duration,Trial)                &     8.82 &      10.56 &     22.17&  $<$ 0.0001    \\ 
re(word)                               &  1734.42 &    2777.00 &      2.03&  $<$ 0.0001    \\ 
re(word,gender)                        &   495.12 &    5544.00 &      0.11&  $<$ 0.0001    \\ 
re(subject, word duration)             &    18.75 &      19.00 &     79.19&  $<$ 0.0001    \\ 
fs(session,subject)                    &   154.36 &     178.00 &    143.62&  $<$ 0.0001    \\ \hline
\end{tabular}
}
\caption{Summary of the model fit to the inverse transformed auditory lexical
decision latencies in the {\tt baldey} megastudy ($\rho = 0.2$).  Factors
received treatment dummy coding, {\tt s()} denotes a thin plate regression
spline, and {\tt te()} a tensor product smooth.  {\tt re(X)} denotes random
intercepts for grouping factor $X$, and {\tt re(X,Y)} specifies random slopes
for $Y$ for grouping factor $X$. {\tt fs()} denotes a penalized factor smooth.
Frequency, word duration, and trial were scaled.}
\label{tab:baldey}
\end{table}

\begin{table}[ht]
\centering
\begin{tabular}{lrrrr} \hline
A. parametric coefficients  & Estimate   & Std. Error & $t$-value    & $p$-value  \\ \hline
  Intercept                 &      -0.94 &       0.04 &      -23.22  & $<$ 0.0001 \\
  gender = male             &       0.24 &       0.06 &        4.14  & $<$ 0.0001 \\
  LemmaFreq                 &      -0.02 &       0.00 &      -12.54  & $<$ 0.0001 \\
  word duration             &       0.06 &       0.01 &        8.75  & $<$ 0.0001 \\
  Trial                     &       0.01 &       0.00 &        7.44  & $<$ 0.0001 \\
  gendermale : LemmaFreq    &       0.00 &       0.00 &        1.38  &     0.1684 \\
  word duration : Trial     &      -0.01 &       0.00 &       -6.44  & $<$ 0.0001 \\ \hline \hline
B. smooth terms             &        edf &     Ref.df &    $F$-value &  $p$-value \\ \hline
  s(word)                   &    1668.12 &    2777.00 &         2.79 & $<$ 0.0001 \\
  s(gender, word)           &     451.77 &    5544.00 &         0.10 & $<$ 0.0001 \\
  s(subject)                &      17.93 &      18.00 &    900901.26 & $<$ 0.0001 \\
  s(word duration, subject) &      18.74 &      19.00 &       121.38 & $<$ 0.0001 \\
  s(session, subject)       &      19.90 &      20.00 &    717027.82 & $<$ 0.0001 \\ \hline
\end{tabular}
\caption{Summary of a model for the {\tt baldey} auditory lexical decision latencies with
only linear effects. Frequency, word duration, and trial were scaled.} 
\label{tab.lineargam}
\end{table}

\newpage

\begin{center}
  {\bf The poems dataset}
\end{center}

\begin{table}[ht]
\centering
{\footnotesize
\begin{tabular}{lrrrr} \hline
                  & Estimate & Std. Error & $t$-value & Pr($>$$|$t$|$) \\ \hline
Intercept         &     6.05 &       0.02 &    347.71 &     $<$ 0.0001 \\ \hline\hline
                  &      edf &     Ref.df &       $F$ &      $p$-value \\ \hline
te(Fre,Trial)     &    10.29 &      11.40 &     80.59 &     $<$ 0.0001 \\ 
re(Poem)          &    81.19 &      87.00 &     19.45 &     $<$ 0.0001 \\ 
fs(Trial, Subject)&  2163.53 &    2934.00 &    329.42 &     $<$ 0.0001 \\ 
re(Subject, Fre)  &   304.11 &     326.00 &     14.88 &     $<$ 0.0001 \\ \hline
\end{tabular}
}
\caption{Summary of the generalized additive mixed model fitted to the 
{\tt poems} data, with $\rho = 0.3$, and a tensor product smooth for 
{\tt Frequency} by {\tt Trial} 
(f{\sc reml} 49300). 
{\tt te(X,Y)} denotes a tensor product smooth, 
{\tt re(X)} random intercepts for grouping factor $X$, 
{\tt re(X,Y)} denotes random slopes for $Y$ by grouping factor $X$, and 
{\tt fs()} denotes a penalized factor smooth.
}
\label{tab:poetry}
\end{table}

\begin{figure}[h]
\centering
\includegraphics[width=0.7\textwidth]{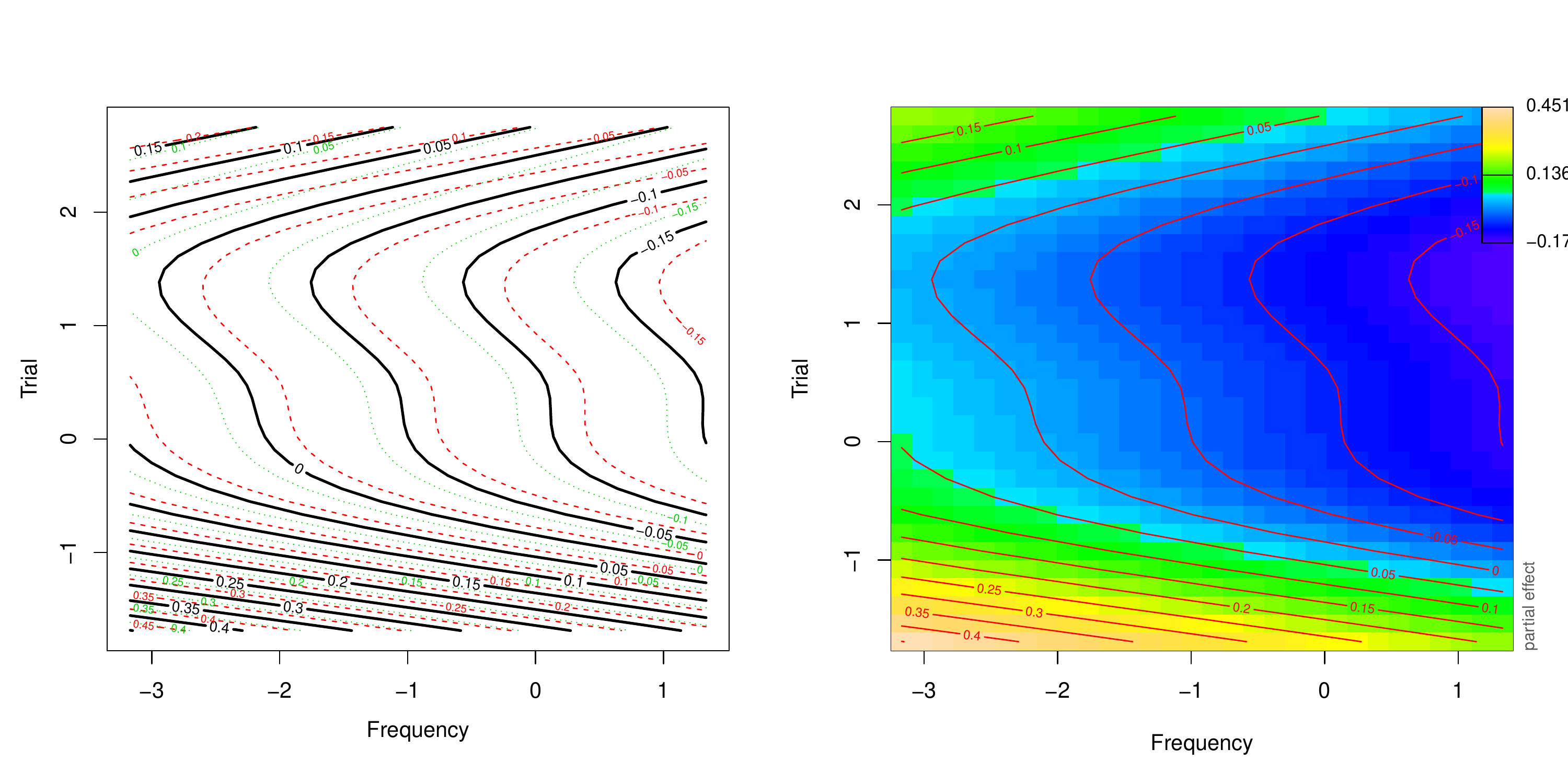}
\caption{Tensor product smooth for the interaction of {\tt Frequency} by {\tt
Trial} in the {\tt poems} data set. In the left panel, green dotted lines
indicate $+1$ standard error contour lines, and red dashed lines $-1$ standard
error contour lines.}
\label{fig:poetryApp}
\end{figure}

\begin{table}[ht]
\centering
{\footnotesize
\begin{tabular}{lrrrr} \hline
                     & Estimate & Std. Error & $t$-value & Pr($>$$|$t$|$) \\ \hline
Intercept            &     6.05 &       0.02 &    347.38 &     $<$ 0.0001 \\ \hline \hline
                     &      edf &     Ref.df &       $F$ &      $p$-value \\ \hline
ti(Fre)              &     1.57 &       1.88 &    247.87 &     $<$ 0.0001 \\ 
ti(TrialSc)          &     3.90 &       3.91 &     90.92 &     $<$ 0.0001 \\ 
ti(Fre,TrialSc)      &     8.05 &      10.19 &      9.93 &     $<$ 0.0001 \\ 
re(Poem)             &    81.18 &      87.00 &     19.41 &     $<$ 0.0001 \\ 
fs(TrialSc, Subject) &  2163.63 &    2934.00 &    323.82 &     $<$ 0.0001 \\ 
re(Subject, Fre)     &   304.08 &     326.00 &     14.87 &     $<$ 0.0001 \\ \hline
\end{tabular}
}
\caption{Summary of a generalized additive mixed model fitted to the {\tt
poems} data, with additive effects of {\tt Frequency}, {\tt Trial}, and their
interaction, and with $\rho = 0.3$ (f{\sc reml}: 49300).  {\tt ti(X,Y)} denotes
an independent tensor product smooth interaction term, and {\tt ti(X)} the
independent main effect of $X$. {\tt re(X)} specifies random intercepts for
grouping factor $X$, and {\tt re(X,Y)} denotes random slopes for $Y$ by
grouping factor $X$.  {\tt fs()} represents a penalized factor smooth, which
absorbs the by-subject random intercepts.
}
\label{tab:ti.poetry}
\end{table}

\begin{table}[ht]
\begin{center}
\begin{tabular}{llrrrr} 
\multicolumn{6}{c}{{\em Random effects}} \\\hline\hline 
 Groups  &Name       &Variance&Std.Dev& Corr &      \\ \hline
 Subject &Intercept  &0.0596&0.2441&      &      \\
         &FreSc      &0.0012&0.0343& -0.61&      \\
         &TrialSc    &0.0096&0.0981&  0.01& 0.04 \\
 Poem    &Intercept  &0.0025&0.0503&      &      \\
 Residual&           &0.0994&0.3153&      &      \\ \hline\hline
\multicolumn{6}{l}{Number of obs: 275996, groups:  Subject, 326; Poem, 87} \\ 
\end{tabular}
\ \\
\vspace*{2\baselineskip}
\begin{tabular}{lrrr} 
\multicolumn{4}{c}{{\em Fixed effects}} \\\hline\hline 
             &  Estimate & Std. Error & t value \\ \hline
Intercept    & 6.0390 &  0.0146 &   414.3 \\
FreSc        &-0.0526 &  0.0020 &   -26.3 \\
TrialSc      &-0.0783 &  0.0055 &   -14.3 \\
FreSc:TrialSc& 0.0039 &  0.0006 &     6.3 \\ \hline\hline
\end{tabular}
\end{center}
\caption{Summary of a linear mixed model fitted to the {\tt poems} data.}
\label{tab:linear.poetry}
\end{table}

\end{document}